\documentclass[sigconf]{acmart}

\usepackage{mathtools}
\usepackage{breakurl}
\usepackage{makecell}
\usepackage{algorithm}
\usepackage{algpseudocode}
\usepackage{booktabs} %For formal tables
\usepackage{enumitem,kantlipsum}
\usepackage{letltxmacro}
\usepackage{multirow,array}
\usepackage{subfig}
\usepackage{scalerel}
\usepackage{listings}	
\usepackage{afterpage,natbib,lipsum}
\usepackage{fancyvrb}
\newcommand\userinput[1]{\textbf{#1}}
\usepackage{url}

\usepackage{float}

\fvset{frame=single,samepage=true,fontfamily=courier,numbersep=2pt,baselinestretch=0.50,fontsize=\small}

\usepackage{tikz}
\usepackage{arydshln}

\usetikzlibrary{backgrounds}

\usepackage{booktabs}

\hypersetup{draft} % maturban should remove If the error "pdfTeX error (ext4): \pdfendlink ended up in different nesting level than \pd" disappears

\begin{document}

\title{Archive Assisted Archival Fixity Verification Framework}

\author{Mohamed Aturban, Sawood Alam, Michael L. Nelson, and Michele C. Weigle}
%\author{a a}
\affiliation{%
  \institution{Old Dominion University}
  %\department{Department of Computer Science}
  \city{Norfolk}
  \state{Virginia}
  \postcode{23529}
  \country{USA}
}
\email{{maturban,salam,mln,mweigle}@cs.odu.edu}
%\email{aa@cs.odu.edu}

% \author{John Doe, Michael L. Nelson, and Michele C. Weigle}
%\author{b b}
%\affiliation{%
%  \institution{Old Dominion University}
  %\department{Department of Computer Science}
%  \city{Norfolk}
%  \state{Virginia}
%  \postcode{23529}
 % \country{USA}
%}
%\email{{jdoe,mweigle,mln}@cs.odu.edu}
%\email{bb@cs.odu.edu}

% \author{John Doe, Michael L. Nelson, and Michele C. Weigle}
%\author{c c}
%\affiliation{%
 % \institution{Old Dominion University}
  %\department{Department of Computer Science}
  %\city{Norfolk}
  %\state{Virginia}
  %\postcode{23529}
  %\country{USA}
%}
%\email{{jdoe,mweigle,mln}@cs.odu.edu}
%\email{cc@cs.odu.edu}

% \author{John Doe, Michael L. Nelson, and Michele C. Weigle}
%\author{d d}
%\affiliation{%
 % \institution{Old Dominion University}
  %\department{Department of Computer Science}
  %\city{Norfolk}
  %\state{Virginia}
  %\postcode{23529}
  %\country{USA}
%}
%\email{{jdoe,mweigle,mln}@cs.odu.edu}
%\email{dd@cs.odu.edu}

% \author{John Doe, Michael L. Nelson, and Michele C. Weigle}
%\author{e e}
%\affiliation{%
  %\institution{Old Dominion University}
  %\department{Department of Computer Science}
  %\city{Norfolk}
  %\state{Virginia}
  %\postcode{23529}
  %\country{USA}
%}
%\email{{jdoe,mweigle,mln}@cs.odu.edu}
%\email{ee@cs.odu.edu}

%\renewcommand{\shortauthors}{M. Aturban}

\begin{abstract}

The number of public and private web archives has increased, and we implicitly trust content delivered by these archives.
% Currently, users can access web archives without the ability to check fixity of content. 
Fixity is checked to ensure an archived resource has remained unaltered since the time it was captured. Some web archives do not allow users to access fixity information and, more importantly, even if fixity information is available, it is provided by the same archive from which the archived resources are requested. In this research, we propose two approaches, namely \emph{Atomic} and \emph{Block}, to establish and check fixity of archived resources. 
%They do not require changes in the current web archiving infrastructure, and it will be built based on well-known web archiving standards, such as the Memento protocol. 
In the \emph{Atomic} approach, the fixity information of each archived web page is stored in a JSON file (or a manifest), and published in a well-known web location (an Archival Fixity server) before it is disseminated to several on-demand web archives. In the \emph{Block} approach, we first batch together fixity information of multiple archived pages in a single binary-searchable file (or a block) before it is published and disseminated to archives. In both approaches, the fixity information is not obtained directly from archives. Instead, we compute the fixity information (e.g., hash values) based on the playback of archived resources. 
One advantage of the \emph{Atomic} approach is the ability to verify fixity of archived pages even with the absence of the Archival Fixity server. The \emph{Block} approach requires pushing fewer resources into archives, and it performs fixity verification faster than the \emph{Atomic} approach. On average, it takes about 1.25X, 4X, and 36X longer to disseminate a manifest to \texttt{perma.cc}, \texttt{archive.org}, and \texttt{webcitation.org}, respectively, than \texttt{archive.is}, while it takes 3.5X longer to disseminate a block to \texttt{archive.org} than \texttt{perma.cc}. The \emph{Block} approach performs 4.46X faster than the \emph{Atomic} approach on verifying the fixity of archived pages.
\end{abstract}

%\begin{CCSXML}
%<ccs2012>
%<concept>
%<concept_id>10002951.10003227.10003392</concept_id>
%<concept_desc>Information systems~Digital libraries and archives</concept_desc>
%<concept_significance>500</concept_significance>
%</concept>
%<concept>
%<concept_id>10002951.10003260</concept_id>
%<concept_desc>Information systems~World Wide Web</concept_desc>
%<concept_significance>500</concept_significance>
%</concept>
%</ccs2012>
%\end{CCSXML}

%\ccsdesc[500]{Information systems~Digital libraries and archives}
%\ccsdesc[500]{Information systems~World Wide Web}

%\keywords{Web Archiving, Fixity, Memento, Security}

\maketitle

\section{Introduction}
Web archives, such as the Internet Archive\footnote{\url{http://archive.org}} (IA) and UK Web Archive\footnote{\url{http://www.webarchive.org.uk/ukwa/}}, have made great efforts to capture and archive the web to allow access to prior states of web resources. We implicitly trust the archived content delivered by such archives, but with the current trend of extended use of other public and private web archives \cite{costa2017,Kim_Nowviskie_Graham_Quon_Alliance_20172}, we should consider the question of validity. For instance, if a web page is archived in 1999 and replayed in 2019, how do we know that it has not been tampered with during those 20 years? One potential solution is to generate a cryptographic hash value on the HTML content of an archived web page, or memento. A memento is an archived version of an original web page \cite{memento:rfc}. Figure \ref{fig:two_commands} shows an example where the \texttt{cURL} command downloads the raw HTML code of the memento 

\begin{center}
\begin{tabular}{l}  
\url{https://web.archive.org/web/20181219102034/https:/}
\\
\url{/2019.jcdl.org/}  
\end{tabular}
\end{center}
{\raggedleft{}a}nd then the hashing function \texttt{sha256sum} generates a SHA-256 hash on this downloaded code.  By running these commands at different times we should always expect to obtain the same hash. 

In the context of web archiving, fixity verifies that archived resources have remained unaltered since the time they were received \cite{fixity_qwens}. The final report of the PREMIS Working Group  \cite{premis2005data} defines information used for fixity as ``information used to verify whether an object has been altered in an undocumented or unauthorized way.''  Web content tampering is a common Internet-related crime in which content is altered by malicious users and activities \cite{kachhawa2014novel}. 

Part of the problem is the lack of standard techniques that users can apply to verify the fixity of web content \cite{%probert2005formal,
 aljawarneh2008design, gao2012iaas}. Jinfang Niu mentioned that none of the web archives declare the reliability of the archived content in their servers, and some archives, such as the Internet Archive, WAX\footnote{\url{wax.lib.harvard.edu/collections/home.do}}, and Government of Canada Web Archive\footnote{\url{www.collectionscanada.gc.ca/webarchives/index-e.html}}, have a disclaimer stating that they are not responsible for the reliability of the archived content they provide \cite{Niu2012}. 

A motivating example, which shows the importance of verifying fixity of mementos, is the story of Joy-Ann Reid, an American cable television host at MSNBC. In December 2017, she apologized for writing several ``insensitive'' LGBT  blog posts nearly a decade ago when she was a morning radio talk show host in Florida \cite{,nelsonplogreid2018,nbcnews2017,nelsonplogreid2018}. In April 2018, Reid, supported by her lawyers, claimed that her blog and/or the archived versions of the blog in the Internet Archive had been compromised and the content was fabricated \cite{mediaite2018}. Even though the Internet Archive denied that their archived pages had been hacked \cite{iadeny2018}, a stronger case could be made if we had an independent service verifying that those archived blog posts had not changed since they were captured by the archive. In this paper, we are introducing two approaches, \emph{Atomic} and \emph{Block}, to make archived web resources verifiable. 

\begin{figure*}
\centering 
%     \begin{minipage}[b]{8.5 cm}
     \begin{Verbatim}[commandchars=\\\{\}]
\userinput{\$ curl} -s https://web.archive.org/web/20181219102034id_/https://2019.jcdl.org/ | \userinput{sha256sum} 

2769b5f71c64794ae76267b3a6385847e202bf0fd9fbcb8e0633427759fac128 -
     \end{Verbatim}
       % \vspace{-1.2em}
     \caption{Commands to generate a hash value of a memento.}
\label{fig:two_commands}
\end{figure*}

In the \emph{Atomic} approach, the fixity information of each archived web page is stored in a single JSON file, or manifest, published on the web, and disseminated to several on-demand web archives. In the \emph{Block} approach, we batch together fixity information, or records, of multiple archived pages to a single binary-searchable file, or block. The block then is published at a well-known web location before disseminating to archives. While we make a chain of blocks, we are not attempting to create yet another Blockchain \cite{bitcoin:academic}. Manifests' chain of blocks are limited in scope as we do not need to worry about \emph{consensus}, \emph{eventual consistency}, or \emph{proof-of-work} because these blocks are generated and published by a central authority (the \emph{Block} approach is described in \ref{sec:block}). In both approaches, the fixity information, such as hash values, is not directly provided by archives (server-side) even though some archives' APIs (e.g., the Internet Archive CDX server \cite{cdxgithub}) allow accessing such information. Alternatively, we decided to calculate the fixity information based on the playback of archived resources (client-side) for two reasons. First, we are not expecting hashes generated and stored in WARC files by archives at crawl time to match those generated on the playback of mementos \cite{turbocplusplus}. Second, if an archive has been compromised then it is likely the corresponding hashes have been also compromised, so we need to have the fixity information stored in independent archives \cite{Maniatis:2003:PPR:945445.945451}.
% In both approaches, the fixity information are not directly provided by archives (e.g., hashes provided by the Internet Archive CDX server \cite{cdxgithub} or hashes in WARC files at \url{https://perma.cc/<Unique-ID-of-a-Memento>?type=warc_download}). Alternatively, we calculate the fixity information, such as hash values, based on the playback of archived resources.} 

This work introduces a basic, yet extensible, format of fixity information in the form of a structured manifest file. However, the main contribution of this paper focuses on the two suggested approaches of disseminating fixity information (or manifests) rather than strength, applicability, extension, scope, or security of the manifest. The framework describes how manifests are published, discovered, and used to verify mementos. The proposed framework does not require any change in the infrastructure of web archives. It is built based on well-known standards, such as the Memento protocol, and works with current archives' APIs. %{\color{blue}The framework is mainly designed to help verifying fixity of mementos with important content. Creating a manifest for every archived page will result in maintaining resources that may never be used for checking fixity (003)}.   
The framework allows for the generation of manifests for selected resources instead of incurring the overhead of creating manifests for all archived resources.

%security (e.g, whether a cryptographic hashing function is vulnerable or not). The framework also describes how manifests are published, discovered, and used to verify mementos ***1***}. {\color{red}The proposed framework does not require any change in the infrastructure of web archives. It is built based on well-known standards, such as the Memento protocol, and works with current archives' APIs (e.g, TimeGate and TimeMap).***6***.}{\color{blue} The framework is mainly designed to help verifying fixity of mementos with important content rather than creating manifests for archived pages that will never be checked for fixity ***3***}. 

%{\color{red}This work defines fixity information, what it consists of, and how it is calculated. However, the main contribution of this paper focus on the two suggested approaches of disseminating fixity information, or manifests, rather than security (e.g, whether a cryptographic hashing function is vulnerable or not). The framework also describes how manifests are published, discovered, and used to verify mementos ***1***}. {\color{red}The proposed framework does not require any change in the infrastructure of web archives. It is built based on well-known standards, such as the Memento protocol, and works with current archives' APIs (e.g, TimeGate and TimeMap).***6***.}{\color{blue} The framework is mainly designed to help verifying fixity of mementos with important content rather than creating manifests for archived pages that will never be checked for fixity ***3***}.   

We show that the size of a manifest represents about 2\% of an actual memento's content, and, on average, it takes about 1.25X, 4X, and 36X longer to disseminate a manifest to \texttt{perma.cc}, the Internet Archive, and WebCite \cite{eysenbach2005going}, respectively, than \texttt{archive.is}, while it takes 3.5X longer to disseminate a block to \texttt{archive.org} than \texttt{perma.cc}. The \emph{Block} approach performs 4.46X faster than the \emph{Atomic} approach on verifying the fixity of archived pages. This paper is an expanded version of a conference paper \cite{aturbanjcdl2019fixity}. 

%\vspace*{-1.5mm}

\section{Background and related work} \label{sec:background-related}
In order to automatically collect portions of the web, web archives employ web crawling software, such as the Internet Archive's Heritrix %mohr2004introduction
 \cite{sigurdhsson2010incremental}. Having a set of seed URIs placed in a queue, Heritrix will start by fetching web pages identified by those URIs, and each time a web page is downloaded, Heritrix writes the page to a WARC file \cite{isowarc}, extracts any URIs from the page, places those discovered URIs in the queue, and repeats the process. 

The crawling process will result in a set of archived pages, or mementos. To provide access to their archived pages, many web archives that use OpenWayback \cite{openwayback}, the open-source implementation of IA's Wayback Machine, to allow users to query the archive by submitting a URI. OpenWayback will replay the content of any selected archived web page in the browser. One of the main tasks of OpenWayback is to ensure that when replaying a web page from an archive, all resources that are used to construct the page (e.g., images, style sheets, and JavaScript files) should be retrieved from the archive, not from the live web. Thus, at the time of replaying the page, OpenWayback will rewrite all links to those resources to point directly to the archive \cite{tofel2007wayback}. In addition to OpenWayback, PyWb \cite{pywb} is another replaying tool, which is used by Perma \cite{perma} and Webrecorder \cite{webrecorder}.
%\subsection{Memento} memento:rfc

Memento \cite{nelson:memento:tr} is an HTTP protocol extension that uses time as a dimension to access the web by relating current web resources to their prior states. The Memento protocol is supported by most public web archives including the Internet Archive. The protocol introduces two HTTP headers for content negotiation. First, Accept-Datetime is an HTTP Request header through which a client can request a prior state of a web resource by providing the preferred datetime (e.g., \emph{Accept-Datetime: Mon, 09 Jan 2017 11:21:57 GMT}).  Second, the Memento-Datetime HTTP Response header is sent by a server to indicate the datetime at which the resource was captured. The Memento protocol also defines the following terminology:
\begin{itemize}[leftmargin=0.3cm]
\item [-] {URI-R - an original resource from the live Web}
\item [-] {URI-M - an archived version (memento) of the original resource at a particular point in time}
\item [-] {URI-T - a resource (TimeMap) that provides a list of mementos (URI-Ms) for a particular original resource}
\item [-] {URI-G - a resource (TimeGate) that supports content negotiation based on datetime to access prior versions of an original resource}
\end{itemize} 

To establish trust in repositories and web archives, different publications and standards have emphasized  the importance of verifying fixity of archived resources. The report Trusted Repositories Audit \& Certification (TRAC) by the Task Force on Archiving of Digital Information introduces criteria for identifying trusted digital repositories \cite{daletrustworthy}.  In addition to the ability to reliably provide access, preserve, and migrate digital resources, digital repositories which include web archives must create preservation metadata that can be used to verify that content is not tampered with or corrupted (fixity) according to sections B2.9 and B4.4. The report recommends that preserved content is stored separately from fixity information, so it is less likely that someone is able to alter both the content and its associated fixity information \cite{daletrustworthy}.  Thus, generating fixity information and using it to ensure that archived resources are valid will help to establish trust in web archives.  Eltgrowth \cite{eltgrowth2009best} outlined several judicial decisions that involve evidence (i.e., archived web pages) taken from the Internet Archive. The author mentions that there is an open question whether to consider an archived web page as a duplicate of the original web page at a particular time in the past. This concern might prevent considering archived web pages as evidence. 

%Eltgrowth mentioned that it is important to find a way to validate content served by a web archive specially for critical resources (e.g., documents used as evidence in courts)\cite{eltgrowth2009best}. 

Different vulnerabilities were discovered in the Internet Archive's Wayback Machine by  Lerner et al. \cite{lerner2017rewriting} and Berlin \cite{johnberlin:thesis}. They are Archive-Escapes, Same-Origin Escapes, Archive-Escapes + Same-Origin Escapes, and Anachronism-Injection. Attackers can leverage these vulnerabilities to modify a user's view at the time when a memento is rendered in a browser. The authors suggested some defenses that could be deployed by either web archives or web publishers to prevent abusing these vulnerabilities. Cushman and Kreymer created a shared repository in May 2017 to describe potential threats in web archives, such as controlling a user's account due to Cross-Site Request Forgery (CSRF) or Cross-Site Scripting (XSS), and archived web resources reaching out to the live web \cite{cushman2017}. The authors provide recommendations on how to avoid such threats. Rosenthal et al. \cite{rosenthali2005}, on the other hand, described several threats against the content of digital preservation systems (e.g., web archives). The authors indicated that designers of archives must be aware of threats, such as media failure, hardware failure, software failure, communication errors, failure of network services, media hardware obsolescence, software obsolescence, operator error, natural disaster, external attack, internal attack, economic failure, and organizational failure. 

Several tools have been developed to generate trusted timestamps. For example, OriginStamp \cite{gipp2015decentralized} allows users to generate a trusted timestamp using blockchain-based networks on any file, plain text, or a hash value. The data is hashed in the user's browser and the resulting hash is sent to OriginStamp's server which then will be added to a list of all hashes submitted by other users. Once per day, OriginStamp generates a single aggregated hash of all received hashes. This aggregated hash is converted to a Bitcoin address that will be a part of a new Bitcoin transaction. The timestamp associated with the transaction is considered a trusted timestamp. A user can verify a timestamp through OriginStamp's API or by visiting their website. Other services, such as Chainpoint (\url{chainpoint.org}) and OpenTimestamps (\url{opentimestamps.org}), are based on the same concept of using blockchain-based networks to timestamp digital documents. Even though users of these services can pass data by value, they are not allowed to submit data by reference (i.e., passing a URI of a web page). In other words, these tools are not directly timestamping web pages. The only exception is a service \cite{gippusingpaperforweb} established by OriginStamp that accepts URIs from users, but the service is no longer available on the live web at 
\begin{center}
\begin{tabular}{l}  
\url{www.isg.uni-konstanz.de/web-time-stamps/}
\end{tabular}
\end{center}
{\raggedleft{}A} number of problems with  blockchain-based networks are descibed by Rosenthal \cite{rosenthal2018blockchain}. He indicates that having a large number of independent nodes in the network is what makes it secure, but this is not the case with many blockchain-based services, such as Ethereum (\url{www.ethereum.org}). 

There are issues related to how web archives preserve and provide access to mementos that make it difficult to generate repeatable fixity information. When serving mementos, web archives often apply some transformation to appropriately replay content in the user's browser. This includes (1) adding archive-specific code to the original content, (2) rewriting links to embedded resources (e.g., images) within an archived page so these resources are retrieved from the archive, not from the live web, and (3) serving content in different file formats like images (or screenshots), ZIP files, and WARC format \cite{kunze2017warc}. Furthermore, issues, such as reconstructing archived web pages, caching, dynamic/randomly-generated content, illustrate how difficult it is to generate repeatable fixity information. Taking into account all of these archive-related issues, it becomes a challenging problem to distinguish between legitimate changes by archives and malicious changes. In our technical report \cite{turbocplusplus} we provide several recommendations of how to generate repeatable fixity information. 

%Shell scripts by Branwen \cite{gwerntimestamping2017} calculate a hash value by considering all resources constructing a web page (e.g., images and scripts) in addition to the HTML content. This seems to be a reasonable solution for timestamping web resources, but without considering the other requirements defined in our previous work \cite{turbocplusplus}, it is difficult to produce a repeatable hash for the same web page over time.

\begin{figure}
  \includegraphics[width=0.47\textwidth]{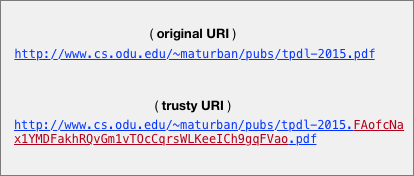}
  %\vspace{-0.9em}
  \caption{An example showing an original URI vs. a trusty URI.}
  %\vspace{-1.9em}
  \label{img:trusty-uri-example}
\end{figure}

Kuhn et al. \cite{kuhn2014trusty} define a trusty URI as a URI that contains a cryptographic hash value of the content it identifies as shown in Figure \ref{img:trusty-uri-example}. %The authors introduced this technique of using trusty URIs to make digital artifacts, especially those related to scholarly publications, immutable, verifiable, and permanent. 

With the assumption that a trusty URI, once created, is linked from other resources or stored by a third party, it becomes possible to detect if the content that the trusty URI identifies has been tampered with or manipulated on the way (e.g., to prevent man-in-the-middle attacks \cite{kuriandistributed}).
%In addition, trusty URIs can verify the content even if it is no longer found at the original URI but still can be retrieved from other locations, such as Google's cache and web archives (e.g., Internet Archive). 
In their second paper \cite{kuhn2015making}, they introduce two different modules to allow creating trusty URIs on different kinds of content. In the module F, the hash is calculated on the byte-level file content, while in the second module R, the hash is calculated on RDF graphs. Even though trusty URIs detect altered documents, there are some limitations. First, a trusty URI is created by an owner of a resource it identifies. Second, trusty URIs can be generated on only two types of content RDF graphs and byte-level content (i.e., no modules introduced for HTML documents). 

% Third, one hash function SHA-256 is used to generate the hash value which might not be suitable for some use cases. Also, there should be alternatives if this hash function becomes vulnerable. 

%Multihash \cite{multihash2017} is a protocol introduced by Juan Benet  to mainly create self identifying hashes for IPFS content \cite{benet2014ipfs}, but the protocol can be actually utilized to identify other content like regular web pages. %The format below is the structure of self described hashes by Multihash \cite{multihash2017}:

%\begin{Verbatim}[commandchars=\\\{\}]
%\{hash function code\}\{digest size\}\{hash value\}
%\end{Verbatim}

%\begin{itemize}
%{\raggedleft{}\textbf{\{hash function code\}}: it is an integer to indicate a hash function used to generate the hash value. For example, $0x11$, $0x12$, and $0x13$ identify $sha1$, $sha2-256$, and $sha2-512$ respectively. The default codes are available on Github\footnote{\url{https://github.com/multiformats/multihash/blob/master/hashtable.csv}} and they are configurable.}
%\item 

%{\raggedleft{}\textbf{\{digest size\}}: it is an integer representing the number of bytes that the hash value consists of.}
%\item 

%{\raggedleft{}\textbf{\{hash value\}}:  the actual hash value with a length of \texttt{\{digest size\}}}
%\end{itemize}

\section{Methodology}

The process of fixity verification of mementos can broadly be described in three phases: 1) generating manifests for mementos, 2) disseminating those manifests into different web archives, and 3) at a later date, generating manifests of the current state and comparing them with their corresponding previously archived versions. We have two approaches of manifest dissemination, namely, \emph{Atomic} and \emph{Block} (as described in Sections \ref{sec:atomic} and \ref{sec:block} respectively). 

\begin{figure*}
\centering
%\userinput
%{\color{red}
\begin{Verbatim}
{   "@context": "http://manifest.ws-dl.cs.odu.edu/",
 
    "created": "Sun, 23 Dec 2018 11:43:55 GMT",
 
    "@id": "http://manifest.ws-dl.cs.odu.edu/manifest/20181223114355/c6ad485819abbe20e37c0632843081710c
            95f94829f59bbe3b6ad3251d93f7d2/https://web.archive.org/web/20181219102034/https://2019.jcdl
            .org/", 
 
    "uri-r": "https://2019.jcdl.org/",

    "uri-m": "https://web.archive.org/web/20181219102034/https://2019.jcdl.org/",
 
    "memento-datetime": "Wed, 19 Dec 2018 10:20:34 GMT",

    "http-headers": {
        
        "Content-Type": "text/html; charset=UTF-8",

        "X-Archive-Orig-date": "Wed, 19 Dec 2018 10:20:36 GMT",

        "X-Archive-Orig-link": "<https://2019.jcdl.org/wp-json/>; rel=\"https://api.w.org/\"",
        
        "Preference-Applied": "original-links, original-content"
    },

    "hash-constructor": "(curl -s '$uri-m' && echo  -n '$Content-Type $X-Archive-Orig-date $X-Archive-O
                          rig-link') | tee >(sha256sum) >(md5sum) >/dev/null | cut -d ' ' -f 1 | paste
                          -d':' <(echo -e 'md5\nsha256') -  | paste -d' ' - -",

    "hash": "md5:969d7aba4c16444a6544bdc39eefe394 sha256:c68a215eb1c3edbf51f565b9a87f49646456369e5179
             1a86106a6667630737a6"      } 
\end{Verbatim}
%}
%\vspace*{-3.2mm}
\caption{A manifest showing fixity information of the memento https://web.archive.org/web/20181219102034/https://2019.jcdl.org/}
\label{json:manifest}
%%\vspace{-1.5em}
\end{figure*}

% The first phase is the generation of manifest files for selected mementos periodically and making them available at well-known URIs. The second phase is the dissemination of those manifests into different web archives by requesting a set of web archives to archive those recently published manifest URIs. The third phase comes on a later date when a memento is asked to be verified by downloading previously archived manifests of the memento and comparing them with a freshly generated manifest of the memento. Optionally, if the freshly generated manifest fails to match with the latest copies in different web archives then it can be pushed/disseminated to archives for future verification.

\subsection{Manifest Generation} \label{sec:manifest}

A manifest (identified by {\texttt{URI-Manif}}) consists of metadata summarizing fixity information of a memento. A manifest can be generated at or after a memento's creation datetime. The proposed structure of a manifest file is illustrated in Figure \ref{json:manifest}, and should have the following properties:\\
{\raggedleft{}\textbf{@context}:} It specifies the URI where names used in the manifest file are defined.

{\raggedleft{}\textbf{created}:} The creation datetime of the manifest. It must be equal to or greater than the memento's creation datetime.

{\raggedleft{}\textbf{URI-R}, \textbf{URI-M}, and \textbf{Memento-Datetime}:} It refers to the URI of an original resource, the URI of a memento, and the datetime when a memento was created, respectively \cite{nelson:memento:tr}.

%{\raggedleft{}\textbf{Memento-Datetime}:} The datetime when a memento was created \cite{nelson:memento:tr}.   
{\raggedleft{}\textbf{@id}:  The URI that identifies a published manifest file (URI-Manif).}

{\raggedleft{}\textbf{http-headers}:  Selected HTTP Response headers of the memento. As proposed by Jones  et al. \cite{rawmemento16}, we insert the \texttt{Preference-Applied} header to specify options used to retrieve the memento. For example, \texttt{Original-Content} refers to the raw memento---accessing unaltered archived content because archives by default return the memento after transforming its content.}

 {\raggedleft{}\textbf{hash-constructor}:  The commands that calculate hashes. The variable \texttt{\$uri-m} is replaced with the \texttt{uri-m} value and the selected headers (e.g., \texttt{\$Content-Type})  are replaced with the corresponding values in the \texttt{http-headers} 
% with their values, \texttt{\$X-Archive-Orig-date}, and \texttt{\$X-Archive-Orig-link} (or any other selected headers)  are replaced with the corresponding values in the \texttt{http-headers}. 
The \texttt{hashes} are generated on both the HTML of a memento and selected response headers, and they are calculated using two different hashing algorithms, \texttt{MD5} and \texttt{SHA256}, so even if the two functions are vulnerable to collision attacks, it becomes difficult for an attacker to make both functions collide at the same time \cite{rosenthal2017sha1dead}.}

 {\raggedleft{}\textbf{hash}:  The hash values calculated based on commands defined in \texttt{hash-constructor}.}

%\begin{figure*}
%  \includegraphics[width=\textwidth]{images/manifest_example}
%  \caption{A manifest shown fixity information of the memento https://web.archive.org/web/20181219102034/https://2019.jcdl.org/.}
%  \label{img:json_manifest}
%\end{figure*}

\subsection{Atomic Dissemination} \label{sec:atomic}

\begin{figure}
\centering
\setlength{\fboxsep}{1pt}%
\fbox{
\includegraphics[width=235px]{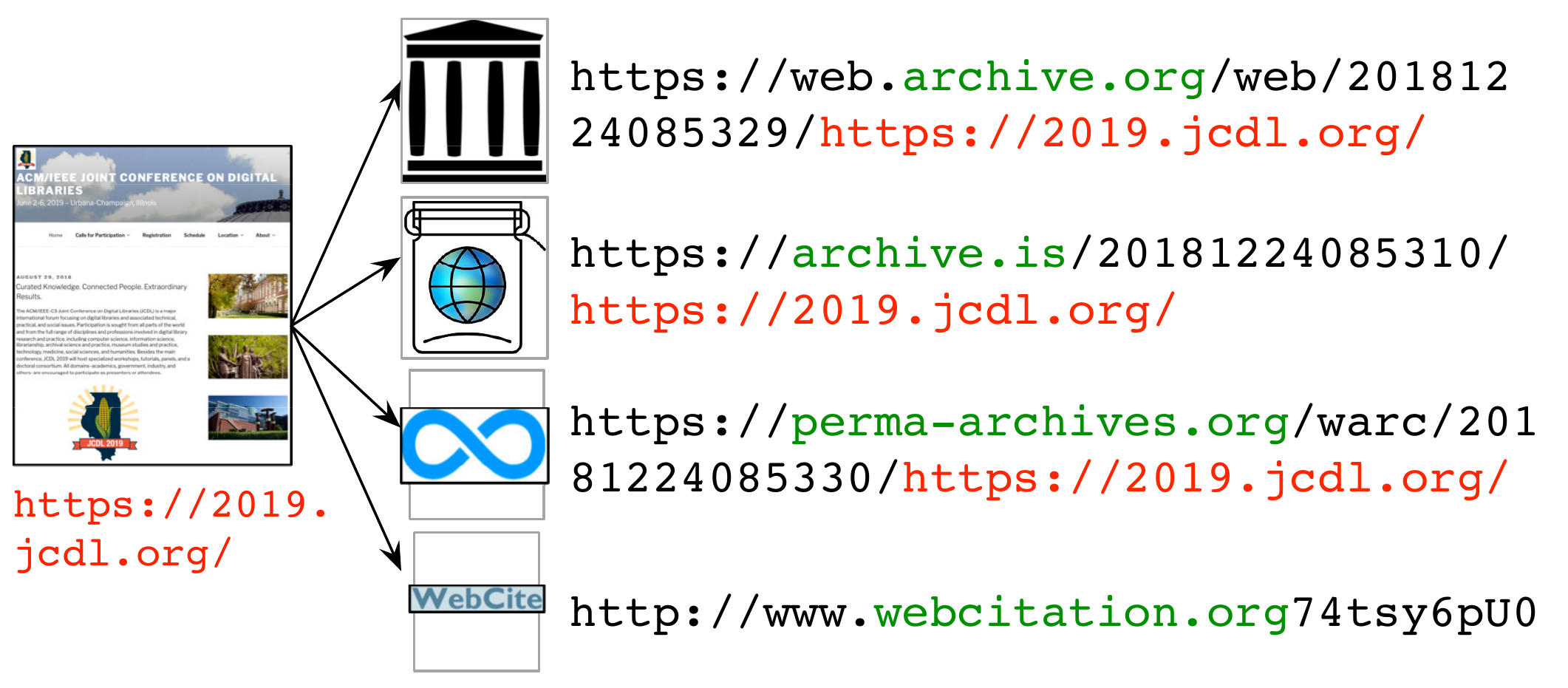}
}
%\vspace*{-3mm}
\caption{A web page is pushed into multiple archives: \texttt{archive.org}, \texttt{archive.is}, \texttt{perma.cc}, and \texttt{webcitation.org}.}
%\vspace*{-6mm}
\label{img:push_ex}
\end{figure}

In the \emph{Atomic} approach, each memento that we are interested in verifying should have at least one corresponding manifest file containing fixity information of the memento. Once generated, the manifest should be published on the web and disseminated to different web archives. The main concept of this approach is to store the fixity information of a memento in differenent archives in addition to the archive in which the memento is preserved. This practice is recommended by the TRAC report \cite{daletrustworthy} where content is maintained  separately from its fixity information. Disseminating manifests can be archived through four steps:
 
\begin{enumerate}%[leftmargin=0.6cm]%[label=(\roman*)]
\item Push a web page into one or more archives. This will create one or more mementos, \texttt{URI-M}.
\item Generate a manifest by computing the fixity information of the memento.
\item Publish the manifest at a well-known location, \texttt{URI-Manif}. 
\item Disseminate the published manifest in multiple archives. This will generate archived manifests, \texttt{URI-M-Manif}.
\end{enumerate}
We briefly describe the steps involved in generating, publishing, and disseminating fixity information of mementos with examples.
%\subsubsection{Generating and storing fixity information}\label{push1step}~\\\\
Figure \ref{img:push_ex} shows the web page \url{https://2019.jcdl.org} pushed into multiple archives, resulting in four mementos. The Python module \texttt{ArchiveNow} can be invoked via the command-line interface or user interface for simultaneously disseminating a web page into on-demand web archives \cite{archivejcdl}.

\begin{figure}
\centering
\setlength{\fboxsep}{0pt}%
\fbox{
\includegraphics[width=235px]{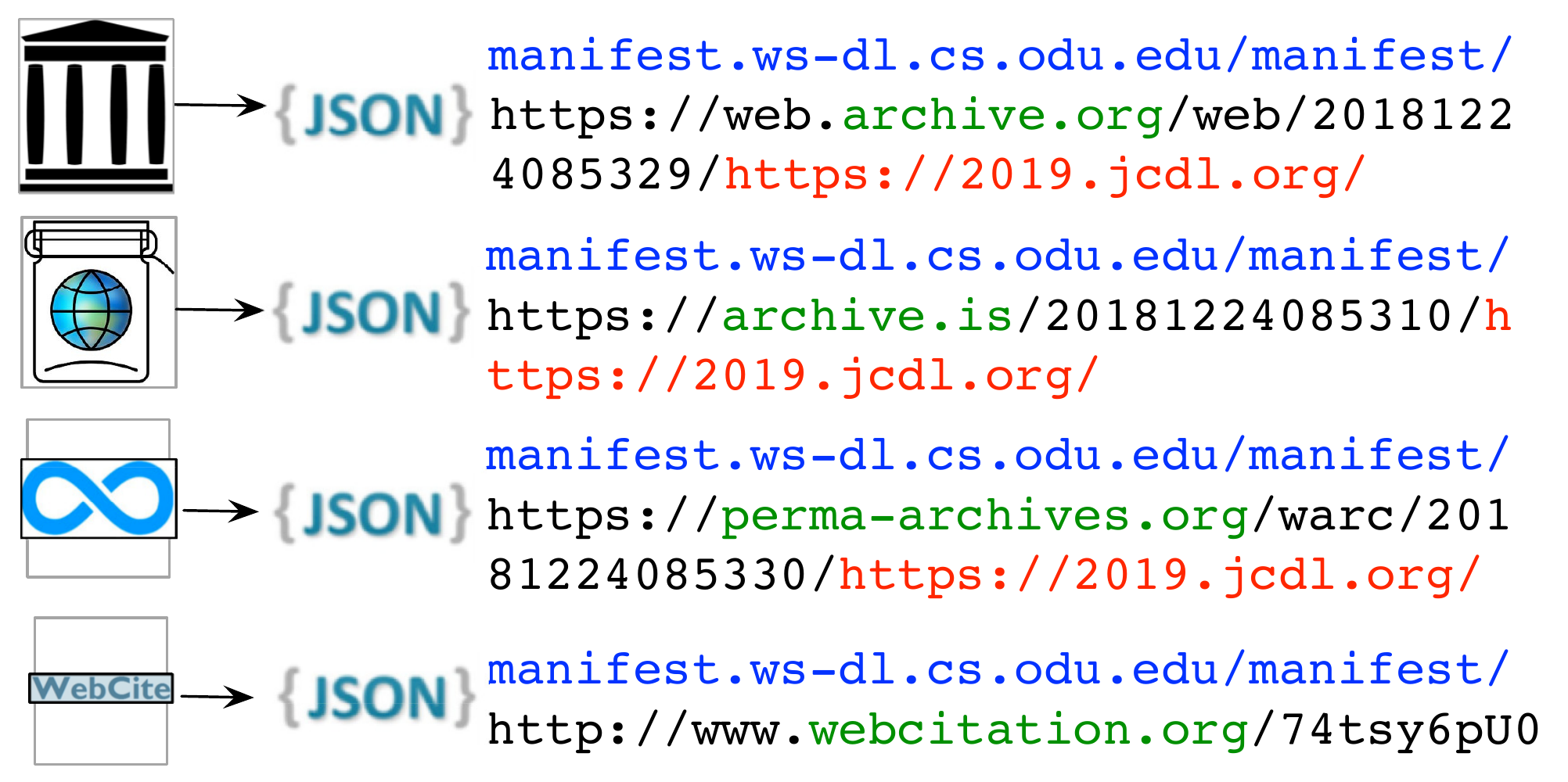}%compute_publish_ex.png}
}
%\vspace*{-3.2mm}
\caption{Compute fixity and publish it on the web}
%\vspace*{-6mm}
\label{img:compute_publish_ex}
\end{figure}

Next, as shown in Figure \ref{img:compute_publish_ex}, for each memento, a manifest is generated and published on the web at the Archival Fixity server,
\begin{center}
\begin{tabular}{l}  
\url{http://manifest.ws-dl.cs.odu.edu}
\end{tabular}
\end{center}
{\raggedleft{}s}o that archives are able to access and capture those manifests. For example, the manifest of the memento 
\begin{center}
\begin{tabular}{l}  
\url{web.archive.org/web/20181224085329/https://2019.}
\\
\url{jcdl.org/}
\end{tabular}
\end{center}
{\raggedleft{}i}s available at the \texttt{URI-Manif}
\begin{center}
\begin{tabular}{l}
{\color{blue}\url{manifest.ws-dl.cs.odu.edu}}\url{/manifest/}{\color{red}\url{https://web.arc}}\\{\color{red}\url{hive.org/web/20181224085329/https://2019.jcdl.org/}}
\end{tabular}
\end{center}

 This URI-Manif is a generic URI, which means if the Archival Fixity server creates another manifest for the same memento (marked in red), the server will publish it using the same generic URI. For this reason, the generic URI must always redirect to the most recent manifest of a memento (i.e., the manifest that is published using a trusty URI), so requesting the manifest's generic URI
\begin{center}
\begin{tabular}{l}  
\url{manifest.ws-dl.cs.odu.edu/manifest/https://web.arc}
\\
\url{hive.org/web/20181224085329/https://2019.jcdl.org/}  
\end{tabular}
\end{center}
{\raggedleft{}w}ill result in ``302 Redirect'' to the trusty URI (Figure \ref{script:manifest-fixity-server-redirect})
\begin{center}
\begin{tabular}{l}  
\url{manifest.ws-dl.cs.odu.edu/manifest/20181224093024/8c}
\\
\url{31ccfbb3a664c9160f98be466b7c9fb9afa80580ab5052001174}
\\
\url{be59c6a73a/https://web.archive.org/web/2018122408532}
\\
\url{9/https://2019.jcdl.org/}  
\end{tabular}
\end{center}

\begin{figure*}
\centering
%\userinput
%{\color{red}
\begin{Verbatim}[commandchars=\\\{\}]
$ curl -sIL https://manifest.ws-dl.cs.odu.edu/manifest/https://web.archive.org/web/20181224085329/https
://2019.jcdl.org/ | egrep -i "(HTTP/|^location:)" 

HTTP/2 302 

location: https://manifest.ws-dl.cs.odu.edu/manifest/20181224093024/8c31ccfbb3a664c9160f98be466b7c9fb9a
fa80580ab5052001174be59c6a73a/https://web.archive.org/web/20181224085329/https://2019.jcdl.org/

HTTP/2 200
\end{Verbatim}
\caption{The manifest identified with the generic URI redirects to the manifest with the trusty URI.}
\label{script:manifest-fixity-server-redirect}

\end{figure*}

Figure \ref{script:timemap-manifest} shows an example of retrieving all mementos (the TimeMap) from the Internet Archive of the URI-Manif: 
\begin{center}
\begin{tabular}{l}  
\url{manifest.ws-dl.cs.odu.edu/manifest/https://web.arc}
\\
\url{hive.org/web/20181224085329/https://2019.jcdl.org/}  
\end{tabular}
\end{center}
{\raggedleft{}As F}igure \ref{script:generic-trusty-redirect} shows, requesting the memento of the manifest (with generic URI) found in the TimeMap %, from Figure  \ref{script:timemap-manifest}, 
results in \texttt{302 Redirect} to the archived manifest (with the trusty URI).

\begin{figure*}
\centering
%\userinput
%{\color{red}
\begin{Verbatim}[commandchars=\\\{\}]
$ curl -i http://web.archive.org/web/\userinput{timemap}/link/https://manifest.ws-dl.cs.odu.edu/manifest/https://web
.archive.org/web/20181224085329/https://2019.jcdl.org/ 


HTTP/1.1 200 OK

Server: nginx/1.15.8

Date: Wed, 01 May 2019 05:38:20 GMT

Content-Type: application/link-format

Transfer-Encoding: chunked

Connection: keep-alive

X-App-Server: wwwb-app38

X-ts: ----

X-location: cdx-p

X-Cache-Key: httpweb.archive.org/web/timemap/link/https://manifest.ws-dl.cs.odu.edu/manifest/https://web
.archive.org/web/20181224085329/https://2019.jcdl.org/US

X-Page-Cache: MISS


<http://manifest.ws-dl.cs.odu.edu/manifest/https://web.archive.org/web/20181224085329/https://2019.jcdl
.org/>; rel="original",

<http://web.archive.org/web/timemap/link/https://manifest.ws-dl.cs.odu.edu/manifest/https://web.archive
.org/web/20181224085329/https://2019.jcdl.org/>; rel="self"; type="application/link-format"; from="Mon,
 24 Dec 2018 09:33:54 GMT",

<http://web.archive.org>; rel="timegate",

<http://web.archive.org/web/20181224093354/http://manifest.ws-dl.cs.odu.edu/manifest/https://web.archive
.org/web/20181224085329/https://2019.jcdl.org/>; rel=\userinput{"first memento"}; datetime="Mon, 24 Dec 2018 09:33:5
4 GMT",
\end{Verbatim}
\caption{Retreiving the TimeMap of a manifest from the Internet Archive. In this example, the TimeMap contains only one memento.}
\label{script:timemap-manifest}

\end{figure*}

\begin{figure*}
\centering
%\userinput
%{\color{red}
\begin{Verbatim}[commandchars=\\\{\}]
$ curl -sIL http://web.archive.org/web/20181224093354/http://manifest.ws-dl.cs.odu.edu/manifest/https://
web.archive.org/web/20181224085329/https://2019.jcdl.org/ | egrep -i "(HTTP/|^location:)"

HTTP/1.1 302 Found

Location: http://web.archive.org/web/20181224093354/http://manifest.ws-dl.cs.odu.edu/manifest/2018122409
3024/8c31ccfbb3a664c9160f98be466b7c9fb9afa80580ab5052001174be59c6a73a/https://web.archive.org/web/201812
24085329/https://2019.jcdl.org/

HTTP/1.1 302 FOUND

Location: http://web.archive.org/web/20181224093355/http://manifest.ws-dl.cs.odu.edu/manifest/2018122409
3024/8c31ccfbb3a664c9160f98be466b7c9fb9afa80580ab5052001174be59c6a73a/https://web.archive.org/web/201812
24085329/https://2019.jcdl.org/

HTTP/1.1 200 OK
\end{Verbatim}
\caption{The archived manifest with the generic URI redirects to the archived manifest with the trusty URI.}
\label{script:generic-trusty-redirect}

\end{figure*}

\begin{figure*}
\centering
\setlength{\fboxsep}{0pt}%
\fbox{
\includegraphics[width=502px]{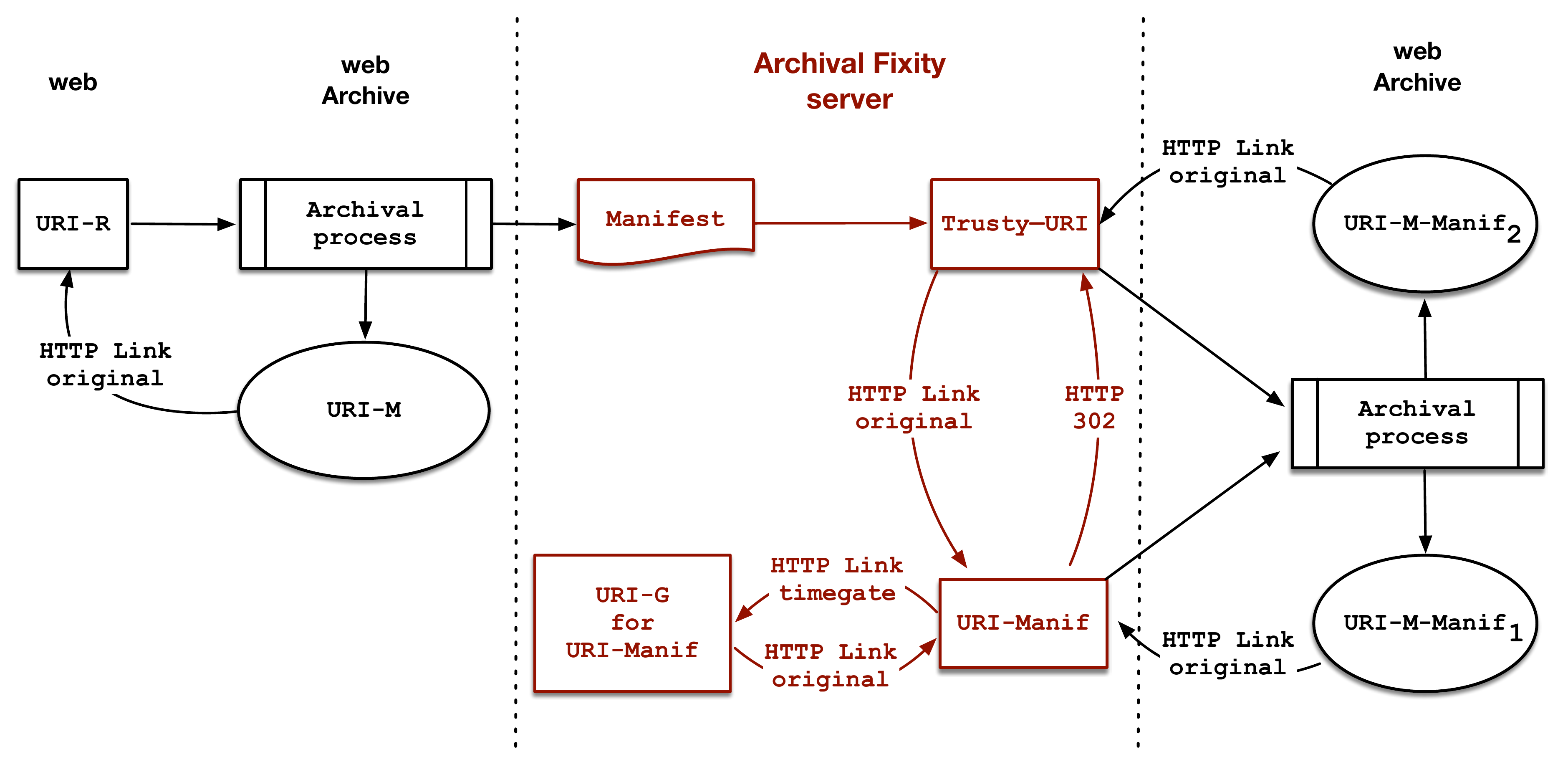}%push_manifest.png}
}
%\vspace*{-4mm}
\caption{The \emph{Atomic} approach. The generic URI (URI-Manif) redirects to the most recent trusty URI, so when the archive captures the generic URI, the archive follows the \texttt{302 Redirect} and captures the trusty URI as well. This figure is a modified version of an original diagram contributed by Herbert Van de Sompel (from DANS).}
%\vspace*{-2mm}
\label{img:atomic_approch_diagram}
\end{figure*}

This \texttt{302 Redirect}  from the generic URI to the trusty URI has two advantages. First, as we described in Section \ref{sec:background-related}, having a trusty URI will help validate the manifest content, as the hash included in the URI is the hash of the content it identifies. Second and more importantly, we can use the generic URI to discover manifests in the Archival Fixity server and archived manifests in the archives. Therefore, even in cases where the Archival Fixity server is unavailable or compromised we still can discover manifests in the archives directly (e.g., using a TimeGate or TimeMap). Figure \ref{img:atomic_approch_diagram} shows how the live web, the archive, and the Archival Fixity server are related in the \emph{Atomic} approach. 

\begin{figure*}
\centering
\setlength{\fboxsep}{0pt}%
\fbox{
\includegraphics[width=502px]{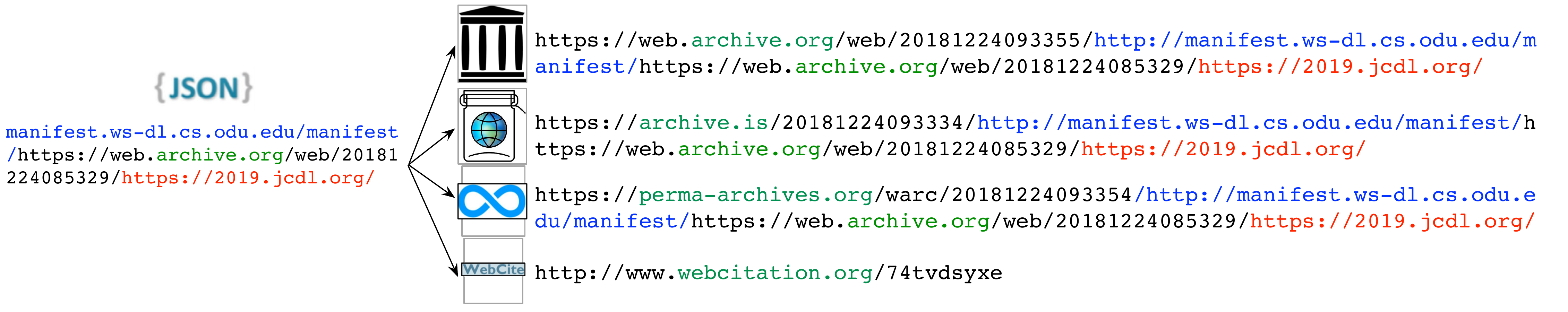}%push_manifest.png}
}
%\vspace*{-4mm}
\caption{Push the fixity information into multiple archives}
%\vspace*{-2mm}
\label{img:push_manifest}
\end{figure*}

Generally, we build trust in the content of memento from the time when fixity information is computed and published. One of the best scenarios is when a manifest is generated at ingest by the archive. In other words, the archive crawls a web page and immediately after that computes and publishes its fixity information.

%, or \texttt{manifest}, of \texttt{URI-M1} and publishes it by submitting this information to \url{manifest.org}. The published manifest is denoted by \texttt{Manifest-URI-1}. The archive pushes \texttt{Manifest-URI-1} into multiple archives. The archived manifest is denoted by \texttt{URI-1} in Figure \ref{img:archive_manifest}. Thus, the archived resource creation date and the creation date of the fixity information should be (close to) the same.  

The final step is to push the published manifest into multiple archives. In the example shown in Figure \ref{img:push_manifest}, the fixity information (or the manifest) of the memento from \texttt{archive.org} is disseminated to the same archive and other three archives including \texttt{archive.is}, \texttt{perma-archives.org}, and \texttt{webcitation.org}.

%Disseminating fixity information in multiple archives has two main benefits. First, it is recommended that fixity information is stored separately from its associated archived content \cite{daletrustworthy}, so it is less likely that someone can alter both the content and its fixity information. The other reason is that having more copies of fixity information distributed among different archives. If an archive becomes unreachable or compromised, we still have other copies. we will explore more places and approaches for storing fixity information, such as blockchain-based networks and InterPlanetary File System (IPFS). In addition, using trusty-URI and Multihash can create more complex URIs that identify fixity information on the web but at the same time those URIs can be used to validate fixity information they identify.

\subsection{Block Dissemination} \label{sec:block}

\begin{figure}
  \includegraphics[width=0.47\textwidth]{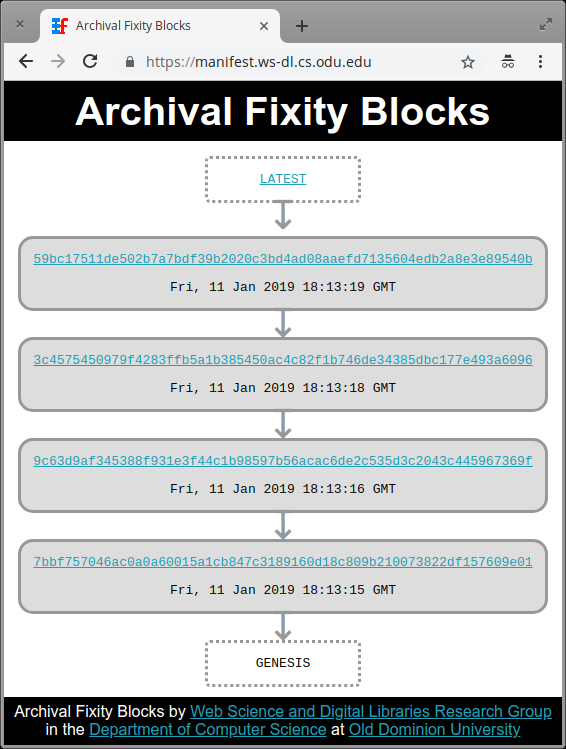}
  %\vspace{-0.9em}
  \caption{The landing page showing a chain of blocks.}
  %\vspace{-1.9em}
  \label{img:blonks-landing-page}
\end{figure}

As opposed to the \emph{Atomic} approach, in the \emph{Block} approach we batch multiple manifests together in a single binary-searchable file along with some additional metadata (using the UKVS file format~\cite{ukvs:sawood,mementomap}), and add the reference of the previously published latest block. Then, we generate the content-addressable identity of the block, compress it, and archive it into multiple web archives by making it available at a well-known content-addressable URI (and allow people to keep local copies anywhere). While we make a chain of blocks, we are not attempting to create yet another Blockchain \cite{bitcoin:academic}.  Manifests' chain of blocks are limited in scope as we do not need to worry about \emph{consensus}, \emph{eventual consistency}, or \emph{proof-of-work} because these blocks are generated and published by a central authority. Linking blocks in a chain using their content-addressable hashes provides tamper-proofing, and enables discovery of previous blocks (starting from the latest or anywhere in the middle of the chain). Additionally, as long as we are depending on an archived page to be available in the archive, we can count on the archived metadata about the page to be available too. 
%Additionally, by leveraging web archives for persistence, we avoid the need of an active network of nodes forever (as is the case with \emph{Cryptocurrency Blockchains}).
Creation and dissemination of manifest blocks is performed in the following steps:
%\vspace*{-1mm}
\begin{enumerate}%[leftmargin=0.5cm]
  \item Identify a set of URI-Ms for their manifests to be included in the same block (a strategically chosen set may improve block compression factor and enable a more efficient lookup for verification later).
  \item Generate their individual manifests in the form of a single-line JSON file (exclude \texttt{@id} field, needed in case of records being placed in a block, and eliminate many common fields that can go in the headers of the block).
  \item Prefix each manifest JSON line with the Sort-friendly URI Reordering Transform (SURT) \cite{surt_format} of the corresponding URI-M.
  \item Write these lines in a UKVS file along with the metadata headers as illustrated in Figure \ref{block-example}.
  \item Add the content-addressable hash of the latest published block in the metadata as the previous block.
  \item Sort the file using \texttt{LC\_ALL=C} locale.
  \item Calculate the content-addressable hash (e.g., SHA256) of this block.
  \item Name the file using its content-addressable hash.
  \item Compress the block file to efficiently archive it.
  \item Publish the compressed block file on a URI that contains its hash.
  \item Make the entrypoint (the well-known URI) redirect to the latest block's URI (as illustrated in Figure~\ref{img:manifest-blocks-api}).
  \item Add \texttt{Link} response header with appropriate links to navigate through the chain of blocks, which is visually illustrated on the landing page as shown in Figure~\ref{img:blonks-landing-page} (a similar approach of creating bidirectional linked list of HTTP messages was used in the HTTPMailbox \cite{httpmailbox:sawood}).
  \item Archive the entrypoint in multiple web archives, which will implicitly archive the latest block as well due to the redirect.
  \item Optionally, for further tamper-proofing post the URI of the newly published block on immutable platforms not controlled by a single authority (e.g., Twitter and GitHub's Gist).
\end{enumerate}

\begin{figure*}
\centering
%\userinput
%{\color{red}
\begin{Verbatim}
$ curl -IL https://manifest.ws-dl.cs.odu.edu/blocks

HTTP/2 302 

content-type: text/html; charset=utf-8

date: Mon, 21 Jan 2019 22:27:14 GMT

location: https://manifest.ws-dl.cs.odu.edu/blocks/59bc17511de502b7a7bdf39b2020c3bd4ad08aaefd7135604e
          db2a8e3e89540b

server: ArchivalFixity/0.1

content-length: 417



HTTP/2 200 

accept-ranges: bytes

cache-control: immutable

content-disposition: attachment; filename="59bc17511de502b7a7bdf39b2020c3bd4ad08aaefd7135604edb2a8e3e8
                     9540b.ukvs.gz"

content-encoding: gzip

content-type: application/ukvs

date: Mon, 21 Jan 2019 22:27:14 GMT

etag: "59bc17511de502b7a7bdf39b2020c3bd4ad08aaefd7135604edb2a8e3e89540b"

expires: Tue, 22 Jan 2019 10:27:14 GMT

last-modified: Fri, 11 Jan 2019 18:19:00 GMT

link: <https://manifest.ws-dl.cs.odu.edu/blocks/59bc17511de502b7a7bdf39b2020c3bd4ad08aaefd7135604edb2a8
      e3e89540b>; rel="self", <https://manifest.ws-dl.cs.odu.edu/blocks/3c4575450979f4283ffb5a1b385450a
      c4c82f1b746de34385dbc177e493a6096>; rel="prev", <https://manifest.ws-dl.cs.odu.edu/blocks/7bbf757
      046ac0a0a60015a1cb847c3189160d18c809b210073822df157609e01>; rel="first", <https://manifest.ws-dl.
      cs.odu.edu/blocks/59bc17511de502b7a7bdf39b2020c3bd4ad08aaefd7135604edb2a8e3e89540b>; rel="last"

server: ArchivalFixity/0.1

content-length: 15227
\end{Verbatim}
%}
%\vspace{-0.5em}
\caption{Blocks Access API}
\label{img:manifest-blocks-api}

\end{figure*}

%\begin{figure*}
%  \includegraphics[width=\textwidth]{images/manifest-blocks-api}
%  \caption{Blocks Access API}
%  \label{img:manifest-blocks-api}
%\end{figure*}

\begin{figure*}
\centering
%\userinput
%{\color{red}
\begin{Verbatim}

!context ["http://oduwsdl.github.io/contexts/fixity"]

!fields {keys: ["surt"]}

!id {uri: "https://manifest.ws-dl.cs.odu.edu/"}

!meta {created_at: "20190111181327"}

!meta {prev_block: "sha256:d4eb1190f9aaae9542fd3ad8a3c4519450cfb00845b632eb2b3f4f098a34144d"}

!meta {type: "FixityBlock"}

org,archive,web)/web/19961022175434/http://search.com/ {<Single line JSON as illustrated in Figure 2>}

org,archive,web)/web/19961219082428/http://sho.com/ {<Single line JSON>}

org,archive,web)/web/19961223174001/http://reference.com/ {<Single line JSON>}

...
\end{Verbatim}
%}
%\vspace{-0.9em}
\caption{A sample Block with metadata headers and records}
\label{block-example}
%\vspace{-1.2em}
\end{figure*}

Although the number of public web archives is increasing \cite{costa2017,Kim_Nowviskie_Graham_Quon_Alliance_20172}, only a few of them support an on-demand web archiving service. However, a small number (greater than one) of independent on-demand archives can suffice for of the purpose of disseminating manifests.
%the number of on-demand web archives that we can disseminate manifests to is relatively small (less than 10 to the best of our knowledge). However, even few independent on-demand archives should be sufficient for our purpose of verifying
The \emph{Block} dissemination approach has a number of advantages over the \emph{Atomic} approach. It requires far fewer network requests to push it to web archives and creates significantly fewer independently published manifest resources to keep track of mementos. By bundling multiple manifests in a single file, it yields a significant compression factor due to the repeated boilerplate content in each manifest file. As web archives die and new ones come to life, these blocks can be replicated and migrated externally to other places efficiently, while in the case of the \emph{Atomic} approach we might lose historical manifests as old web archives die without donating their holdings to live archives. Moreover, these blocks are more tamper-proof than atomic manifests due to chaining. On the other hand, the \emph{Block} approach has the disadvantage of shifting the burden of lookup of a specific record in the entire chain of blocks to the user or a service that provides verification. While individual blocks are binary searchable for fast lookup, as the number of blocks increases, one has to scan through all of them. However, this can easily be solved by scanning the entire chain once and creating a search index over the SURT field.

\subsection{Verifying Fixity of Mementos}  \label{sec:verifying}

Verifying the fixity of a memento in both the \emph{Atomic} and \emph{Block} approaches can be achieved through three common steps: 
\begin{enumerate}[label=(\alph*)]
\item For the given memento, discover one or more manifests \texttt{URI-Manif}. In the \emph{Atomic} approach, this step requires also discovering  archived copies  \texttt{URI-M-Manif} of the manifest.  
\item Recompute current fixity information of the memento.
\item Compare current fixity information with discovered manifests.
\end{enumerate}

In the \emph{Atomic} approach, we can discover a manifest of a given memento through the Archival Fixity server. Which manifest is returned depends on the server's API. For example, the server may respond with the closest manifest to the memento's creation date or return the manifest that is closest to a given datetime (i.e., via a TimeGate). Once a manifest is discovered,  we may use TimeGates and/or TimeMaps to retrieve its archived copies available in web archives. Again, it is possible to discover archived manifests using the generic URI even without the Archival Fixity server being involved. Next, we compute current fixity information by generating a new manifest for the given memento. Then, we compare current hash values in the new manifest with the hashes in the discovered archived manifests. In this compression step, we should only consider independent copies of the manifest. For example, if an archived manifest is delivered from the same archive where the memento is from, then this copy of the manifest should not be considered independent. In other cases, two manifests might be discovered in two cooperating archives (e.g., we know Archive-It.org is a service established by the Internet Archive). 

In case of the \emph{Block} approach, the fixity verification server (or any equivalent tool) needs to have access to all the blocks, either over HTTP (e.g., from a web archive) or stored locally. These blocks are then scanned for one or more matching records for the given \emph{URI-M}. Corresponding single-line JSON entries (as shown in Figure~\ref{block-example}) are extracted as historical fixity records for comparison. Remaining steps for creating current fixity information, comparing with the historical records, and generating the response summary are the same as in the case of \emph{Atomic} approach. 

While due to the immutable nature of blocks we can only have back references, creating a single linked list pointing from the most recent blocks to the older ones, with the help of some external metadata our archival fixity block server provides bidirectional navigational links for easy navigation along the chain back and forth (as illustrated in Figure~\ref{img:manifest-blocks-api} with \texttt{first}, \texttt{last}, \texttt{prev}, and \texttt{next} link relations in the \texttt{link} header). The content of these blocks is sorted that enables fast lookup in each block using binary search, but the chain of blocks has to be scanned linearly, which can decrease the throughput as the number of blocks increases. To deal with this issue, one can create an inverted index of existing blocks, treating \emph{URI-Ms} as the keyword and blocks as the document. Additionally, the chain of blocks is in chronological order, which makes it easy to create a lightweight skip index to identify segments of the chain that were created around certain points of time in the past. Creating large blocks with a slowly growing chain will be more efficient than a rapidly growing chain of small blocks. However, an optimal block size can be decided based on how long one is willing to wait for enough records to be available for a new block creation and on the largest size of a single block that can easily be stored in web archives. Creating blocks with strategically grouped URI-Ms (e.g., mementos of nearby \texttt{datetime} values, URI-Rs from a set of domains, or URI-Ms from a set of archives) can also improve the efficiency of lookup (or indexing).

\section{Evaluation}
We conducted a study on 1,000 mementos from the Internet Archive which are a subset of a larger set of URI-Ms involved in a different research project \cite{maturabn2018computefixity}. We did not take the size of mementos into consideration (i.e., the number of embedded resources, such as images and JavaScript/CSS files) because fixity in this paper is computed based on only the returned raw HTML content of the base file. The main reason for choosing a small set, only 1,000 URI-Ms, is because the study requires pushing at least 12 manifests for each memento in multiple archives. Sending too many archiving requests to archives might result in technical issues, such as blocking IP addresses. For example, \texttt{webcitation.org} responded with ``WebCite has flagged your IP address for suspicious activity'' after making 100 requests, but the issue was resolved after contacting the archive. Perma.cc on the other hand allows users to freely submit a maximum of 10 URIs for preserving per month. Fortunately, the archive supported us for this study by increasing this limit, so we were able to disseminate more manifests to the archive. Part of evaluation is measuring the time it takes to generate, disseminate, and verify manifests in the \emph{Atomic} and \emph{Block} approaches. In addition, we want to compare the size of files created in both approaches and whether all mementos are going to be verified successfully.  

We wrote Python scripts  \cite{archivalfixity_maturban_sawood} for performing different functions:

\vspace{2mm}

{\raggedleft{}\textbf{generate\_atomic()}:} Accepts a URI-M and returns the filename of a JSON file containing the fixity information of the memento. We generated 3,000 manifests. The main purpose of generating three manifests for each memento is because we are interested in reporting the average time for generating a manifest for each memento. Figure \ref{script:generate_manifest_ex} shows an example of generating a manifest of the memento
\begin{center}
\begin{tabular}{l}  
\url{https://web.archive.org/web/20051123211159/http://}
\\
\url{www.whitehouse.org}
\end{tabular}
\end{center}
{\raggedleft{}T}he resulting JSON file contains the fixity information including the hash calculated on the returned HTML of the memento.
\begin{figure*}
\centering 
     %\begin{minipage}[b]{8.5 cm}
     \begin{Verbatim}[commandchars=\\\{\}]
$ python fixity.py \userinput{generate_atomic} https://web.archive.org/web/20051123211159/http://www.whitehouse.org 


43423e8ad464461ec196c21033451c07b71b1ec4fbd3be013e3235093abac56b.json
\end{Verbatim}        
 \caption{An example of generating a manifest of a memento.}
\label{script:generate_manifest_ex}
\end{figure*}

\vspace{2mm}

{\raggedleft{}\textbf{publish\_atomic()}:} Submits a given JSON to the Archival Fixity server at 

\begin{tabular}{l}  
\url{https://manifest.ws-dl.cs.odu.edu}
\end{tabular}

{\raggedleft{}T}he server will insert \texttt{@id} and \texttt{created} metadata before publishing the new manifest on the web. Figure \ref{script:publish_manifest_ex} shows an example of publishing the manifest file generated previously (in Figure \ref{script:generate_manifest_ex}). It returns the generic URI of the manifest \texttt{URI-Manif} and the trusty URI.
\begin{figure*}
\centering 
     %\begin{minipage}[b]{8.5 cm}
     \begin{Verbatim}[commandchars=\\\{\}]
$ python fixity.py \userinput{publish_atomic} 43423e8ad464461ec196c21033451c07b71b1ec4fbd3be013e3235093abac56b.json

http://manifest.ws-dl.cs.odu.edu/manifest/https://web.archive.org/web/20051123211159/http://www.whiteho
use.org

http://manifest.ws-dl.cs.odu.edu/manifest/20181212013507/43423e8ad464461ec196c21033451c07b71b1ec4fbd3be
013e3235093abac56b/https://web.archive.org/web/20051123211159/http://www.whitehouse.org
\end{Verbatim}        
 \caption{An example of publishing a manifest at the Archival Fixity server.}
\label{script:publish_manifest_ex}
\end{figure*}
\vspace{2mm}

{\raggedleft{}\textbf{disseminate\_atomic()}:} Pushes a published manifest into different archives using ArchiveNow. In our study, we used \texttt{archive.org}, \texttt{archive.is}, \texttt{perma.cc}, and \texttt{webcitation.org} resulting in creating 12,000  archived manifests (i.e., 3,000 URI-M-Manif in each archive). We used the Generic URI to push manifest into archives. Again, this URI always redirects to the trusty URI. If archives consider a ``302 Redirect'' as a separate resource, then the total number of archived resources created in the four archives was 24,000. Figure \ref{script:disseminate_manifest_ex} shows an example of disseminating to four archives the manifest
\begin{center}
\begin{tabular}{l}  
\url{http://manifest.ws-dl.cs.odu.edu/manifest/https://}
\\
\url{web.archive.org/web/20051123211159/http://www.whit}
\\
\url{ehouse.org}
\end{tabular}
\end{center}

\begin{figure*}
\centering 
     %\begin{minipage}[b]{8.5 cm}
     \begin{Verbatim}[commandchars=\\\{\}]
$ python fixity.py \userinput{disseminate_atomic} http://manifest.ws-dl.cs.odu.edu/manifest/https://web.archive.org
/web/20051123211159/http://www.whitehouse.org

http://archive.is/egyVY

https://perma.cc/VMQ3-E45U

http://www.webcitation.org/74bAo5hJ4

https://web.archive.org/web/20181212013856/http://manifest.ws-dl.cs.odu.edu/manifest/20181212013507/434
23e8ad464461ec196c21033451c07b71b1ec4fbd3be013e3235093abac56b/https://web.archive.org/web/2005112321115
9/http://www.whitehouse.org

\end{Verbatim} 
 \caption{An example of disseminating a manifest to four archives.}
\label{script:disseminate_manifest_ex}
\end{figure*}
\vspace{2mm}

{\raggedleft{}\textbf{verify\_atomic()}:} Accepts a URI-M, It discovers a manifest closest to the memento's creation datetime. In addition, the function discovers archived copies of the manifest in the four archives using TimeGates and TimeMaps. Then, it computes current fixity information using \texttt{generate\_atomic()}. Finally, it compares current fixity information with the discovered manifests and their archived copies. As a result, for each URI-M, the function returns either ``Verified'' or ``Failed'' with other information, such as hash values, URI-Manifs, and URI-M-Manifs. Figure \ref{script:verify_manifest_ex} shows an example of verifying the fixity of the memento 
\begin{center}
\begin{tabular}{l}  
\url{https://web.archive.org/web/20051123211159/http://}
\\
\url{www.whitehouse.org.}
\end{tabular}
\end{center}

\begin{figure*}
\centering 
     %\begin{minipage}[b]{8.5 cm}
     \begin{Verbatim}[commandchars=\\\{\}]
$ python fixity.py \userinput{generate_atomic} https://web.archive.org/web/20051123211159/http://www.whitehouse.org

0c6c4f7435a79e9a756fb892dc602f9cd1e71a7f74b0346d999b6c1834c703a4.json


$ python fixity.py \userinput{verify_atomic} 0c6c4f7435a79e9a756fb892dc602f9cd1e71a7f74b0346d999b6c1834c703a4.json

Current hash: md5:c4cf615c62a14df9ac2d610873794555 sha256:e5832410226e1e9637eb9fb7c97cf2065851fb106b925
1412e2821965e02305c

5 matched manifests:

        http://archive.is/egyVY
        
        https://perma.cc/VMQ3-E45U
        
	http://www.webcitation.org/74bAo5hJ4
	
        http://manifest.ws-dl.cs.odu.edu/manifest/20181212013507/43423e8ad464461ec196c21033451c07b71b1e
        c4fbd3be013e3235093abac56b/https://web.archive.org/web/20051123211159/http://www.whitehouse.org
        
	https://web.archive.org/web/20181212013856/http://manifest.ws-dl.cs.odu.edu/manifest/2018121201
	3507/43423e8ad464461ec196c21033451c07b71b1ec4fbd3be013e3235093abac56b/https://web.archive.org/w
	eb/20051123211159/http://www.whitehouse.org
	

0 mismatched manifests:

\end{Verbatim} 
 \caption{An example of verifying the fixity of the memento \url{https://web.archive.org/web/20051123211159/http://www.whitehouse.org}. The current fixity information should be generated first. Then, the function \texttt{verify\_atomic()} finds a published manifest in the Archival Fixity server and its archived versions in web archives. Finally, the function compares current fixity information with the fixity information in the discovered  manifest and its archived captures.}
\label{script:verify_manifest_ex}
\end{figure*}
\vspace{2mm}

{\raggedleft{}\textbf{generate\_block()}:} Accepts multiple JSON files. It generates one or more blocks depending on the selected block size. In this study, we set it to 100 manifests per block, so the total number of generated blocks was 10. The example in Figure \ref{script:generate_block_ex_0} shows the output of the shell script \texttt{generate\_block.sh}, which uses the Python function \texttt{generate\_block()} to generate ten 100-record blocks. Figure \ref{script:generate_block_ex} shows only four records (out of 100) of block 1. The four records have the fixity information of the following mementos:
\begin{center}
\begin{tabular}{l}  
{\Large \textbullet} \texttt{https://web.archive.org/web/19961022175434/http:}
\\
\texttt{//www.search.com:80}  
\\
{\Large \textbullet} \texttt{https://web.archive.org/web/19961023041557/http:}
\\
\texttt{//www.aaas.org:80/}  
\\
{\Large \textbullet} \texttt{https://web.archive.org/web/19961219082428/http:}
\\
\texttt{//www.sho.com:80/}
\\
{\Large \textbullet} \texttt{https://web.archive.org/web/19961223174001/http:}
\\
\texttt{//www2.reference.com:80/}
\end{tabular}
\end{center}

\begin{figure*}
\centering 
\begin{Verbatim}

$./generate_blocks.sh

Input:      urims.txt
Output Dir: ./blocks/100
Block Size: 100
Num Blocks: 10
======================
[1548318100009] Generating block 1
[1548318100348] Saving 87606 bytes to ./blocks/100/20190124082140-00000000000000000000000000000000000000
00000000000000000000000000-dfbbe3600d5fe4e51c895db94cb9e9cfd0eb04716d9e4be6e63cf8ac3f3e9233.ukvs
[1548318100350] Compressing block to ./blocks/100/20190124082140-000000000000000000000000000000000000000
0000000000000000000000000-dfbbe3600d5fe4e51c895db94cb9e9cfd0eb04716d9e4be6e63cf8ac3f3e9233.ukvs.gz
[1548318100356] Finished creating block 1 of size 15174 bytes in 347 milliseconds
======================
[1548318101360] Generating block 2
[1548318101686] Saving 82971 bytes to ./blocks/100/20190124082141-dfbbe3600d5fe4e51c895db94cb9e9cfd0eb04
716d9e4be6e63cf8ac3f3e9233-861f2b2e872125f31a61bed8141f1c8be04c48ebbebb2a49b4fdf2d9d6999f77.ukvs
[1548318101688] Compressing block to ./blocks/100/20190124082141-dfbbe3600d5fe4e51c895db94cb9e9cfd0eb047
16d9e4be6e63cf8ac3f3e9233-861f2b2e872125f31a61bed8141f1c8be04c48ebbebb2a49b4fdf2d9d6999f77.ukvs.gz
[1548318101693] Finished creating block 2 of size 14233 bytes in 333 milliseconds
======================
[1548318102695] Generating block 3
[1548318103017] Saving 84359 bytes to ./blocks/100/20190124082143-861f2b2e872125f31a61bed8141f1c8be04c48
ebbebb2a49b4fdf2d9d6999f77-213c79bda7483d87609287142b86bc8d6b8c66306662236455507be046b0caf2.ukvs
[1548318103019] Compressing block to ./blocks/100/20190124082143-861f2b2e872125f31a61bed8141f1c8be04c48e
bbebb2a49b4fdf2d9d6999f77-213c79bda7483d87609287142b86bc8d6b8c66306662236455507be046b0caf2.ukvs.gz
[1548318103025] Finished creating block 3 of size 14516 bytes in 330 milliseconds
======================
[1548318104030] Generating block 4
[1548318104355] Saving 92165 bytes to ./blocks/100/20190124082144-213c79bda7483d87609287142b86bc8d6b8c66
306662236455507be046b0caf2-4fa056c839656babdd8c8428df006590d7f48a4fbcd7df2d76d3d77110eba056.ukvs
[1548318104357] Compressing block to ./blocks/100/20190124082144-213c79bda7483d87609287142b86bc8d6b8c663
06662236455507be046b0caf2-4fa056c839656babdd8c8428df006590d7f48a4fbcd7df2d76d3d77110eba056.ukvs.gz
[1548318104362] Finished creating block 4 of size 16321 bytes in 332 milliseconds
======================
[1548318105366] Generating block 5
[1548318105689] Saving 92646 bytes to ./blocks/100/20190124082145-4fa056c839656babdd8c8428df006590d7f48a
4fbcd7df2d76d3d77110eba056-70524610a0b3736de2c9b7ea9988ed1cedbcf3098c0f35d2cfed5f89d3193a45.ukvs
[1548318105691] Compressing block to ./blocks/100/20190124082145-4fa056c839656babdd8c8428df006590d7f48a4
fbcd7df2d76d3d77110eba056-70524610a0b3736de2c9b7ea9988ed1cedbcf3098c0f35d2cfed5f89d3193a45.ukvs.gz
[1548318105697] Finished creating block 5 of size 16136 bytes in 331 milliseconds
======================
[1548318106701] Generating block 6
[1548318107032] Saving 90681 bytes to ./blocks/100/20190124082147-70524610a0b3736de2c9b7ea9988ed1cedbcf3
098c0f35d2cfed5f89d3193a45-8d40976dcce88bbc9c5907618e2f95beaa7484426fff509b9fa42a92719edae3.ukvs
[1548318107033] Compressing block to ./blocks/100/20190124082147-70524610a0b3736de2c9b7ea9988ed1cedbcf30
98c0f35d2cfed5f89d3193a45-8d40976dcce88bbc9c5907618e2f95beaa7484426fff509b9fa42a92719edae3.ukvs.gz
[1548318107038] Finished creating block 6 of size 15756 bytes in 337 milliseconds
======================
[1548318108043] Generating block 7
[1548318108379] Saving 89160 bytes to ./blocks/100/20190124082148-8d40976dcce88bbc9c5907618e2f95beaa7484
426fff509b9fa42a92719edae3-e0ef8e7677778fc430d6142b87204f0510e81007c4019747e32e76851e30f657.ukvs
[1548318108381] Compressing block to ./blocks/100/20190124082148-8d40976dcce88bbc9c5907618e2f95beaa74844
26fff509b9fa42a92719edae3-e0ef8e7677778fc430d6142b87204f0510e81007c4019747e32e76851e30f657.ukvs.gz
[1548318108387] Finished creating block 7 of size 15604 bytes in 344 milliseconds
======================
[1548318109391] Generating block 8
[1548318109718] Saving 89228 bytes to ./blocks/100/20190124082149-e0ef8e7677778fc430d6142b87204f0510e810
07c4019747e32e76851e30f657-59851a4cb1e29b9e178c24bfb31a2043762c554d34ac1a34e3cf0880ee7d87be.ukvs
[1548318109720] Compressing block to ./blocks/100/20190124082149-e0ef8e7677778fc430d6142b87204f0510e8100
7c4019747e32e76851e30f657-59851a4cb1e29b9e178c24bfb31a2043762c554d34ac1a34e3cf0880ee7d87be.ukvs.gz
[1548318109725] Finished creating block 8 of size 15550 bytes in 334 milliseconds
======================
[1548318110729] Generating block 9
[1548318111066] Saving 87937 bytes to ./blocks/100/20190124082151-59851a4cb1e29b9e178c24bfb31a2043762c55
4d34ac1a34e3cf0880ee7d87be-2b01231e1a07d92bc5e5aa3b4b3a76e3672384df69013c3537a2cd7505cd23d0.ukvs
[1548318111068] Compressing block to ./blocks/100/20190124082151-59851a4cb1e29b9e178c24bfb31a2043762c554
d34ac1a34e3cf0880ee7d87be-2b01231e1a07d92bc5e5aa3b4b3a76e3672384df69013c3537a2cd7505cd23d0.ukvs.gz
[1548318111074] Finished creating block 9 of size 15354 bytes in 345 milliseconds
======================
[1548318112078] Generating block 10
[1548318112400] Saving 88688 bytes to ./blocks/100/20190124082152-2b01231e1a07d92bc5e5aa3b4b3a76e3672384
df69013c3537a2cd7505cd23d0-5da03e339e52bafcc82b64c1636adff474a94df46a057e9356e74f70eba8b26f.ukvs
[1548318112402] Compressing block to ./blocks/100/20190124082152-2b01231e1a07d92bc5e5aa3b4b3a76e3672384d
f69013c3537a2cd7505cd23d0-5da03e339e52bafcc82b64c1636adff474a94df46a057e9356e74f70eba8b26f.ukvs.gz
[1548318112408] Finished creating block 10 of size 15223 bytes in 330 milliseconds
======================
\end{Verbatim} 
 \caption{The shell script uses the Python function \texttt{generate\_block()} to generate ten 100-record blocks.}
\label{script:generate_block_ex_0}
\end{figure*}
\vspace{2mm}

\begin{figure*}
\centering 
\begin{Verbatim}

!context ["http://oduwsdl.github.io/contexts/fixity"]
!fields {keys: ["surt"]}
!id {uri: "https://manifest.ws-dl.cs.odu.edu/"}
!meta {created_at: "20190111181327"}
!meta {prev_block: "sha256:d4eb1190f9aaae9542fd3ad8a3c4519450cfb00845b632eb2b3f4f098a34144d"}
!meta {type: "FixityBlock"}

org,archive,web)/web/19961022175434/http:/www.search.com:80 {"created": "Wed, 12 Dec 2018 08:54:26 GMT",
"hash": "md5:d6332bc887d295fa02d0e36fe4e2991b sha256:967a9261bb201830224cdb40740a7bdd2b05aa6287b6bbfa3c
f1e8bca9a4e62d", "hash-constructor": "(curl -s '$uri-m' && echo -n '$Content-Type $X-Archive-Orig-date') 
| tee >(sha256sum) >(md5sum) >/dev/null | cut -d ' ' -f 1 | paste -d':' <(echo -e 'md5\nsha256') - | 
paste -d' ' - -", "http-headers": {"Content-Type": "text/html","Preference-Applied": "original-links, 
original-content", "X-Archive-Orig-date": "Tuesday, 22-Oct-96 17:54:34 GMT"}, "memento-datetime": "Tue, 
22 Oct 1996 17:54:34 GMT", "uri-m": "https://web.archive.org/web/19961022175434/http://www.search.com:80
/", "uri-r": "http://www.search.com:80/"}

org,archive,web)/web/19961023041557/http:/www.aaas.org:80 {"created": "Wed, 12 Dec 2018 09:02:44 GMT", 
"hash": "md5:999a5d5012b8072565642e5b55507a3b sha256:02c777b5c3961c05602bd876243db130f84cc66895c3a4cef7
8d52afa4beac4b", "hash-constructor": "(curl -s '$uri-m' && echo -n '$Content-Type $X-Archive-Orig-date 
$X-Archive-Orig-last-modified') | tee >(sha256sum) >(md5sum) >/dev/null | cut -d ' ' -f 1 | paste -d':' 
<(echo -e 'md5\nsha256') - | paste -d' ' - -", "http-headers": {"Content-Type": "text/html", "Preference
-Applied": "original-links, original-content", "X-Archive-Orig-date": "Wednesday, 23-Oct-96 04:15:57 
GMT", "X-Archive-Orig-last-modified": "Friday, 18-Oct-96 14:07:23 GMT"}, "memento-datetime": "Wed, 23 
Oct 1996 04:15:57 GMT", "uri-m": "https://web.archive.org/web/19961023041557/http://www.aaas.org:80/", 
"uri-r": "http://www.aaas.org:80/"}

org,archive,web)/web/19961219082428/http:/www.sho.com:80 {"created": "Wed, 12 Dec 2018 08:58:35 GMT", 
"hash": "md5:07c132d54ce420f1132affec24624cdf sha256:41186b26814213f587493da45d2148db962130fd47a8616c97
f9d4335e5e217a", "hash-constructor": "(curl -s '$uri-m' && echo -n '$Content-Type $X-Archive-Orig-date 
$X-Archive-Orig-last-modified') | tee >(sha256sum) >(md5sum) >/dev/null | cut -d ' ' -f 1 | paste -d':' 
<(echo -e 'md5\nsha256') - | paste -d' ' - -", "http-headers": {"Content-Type": "text/html", "Preference
-Applied": "original-links, original-content", "X-Archive-Orig-date": "Thu, 19 Dec 1996 08:24:34 GMT",
"X-Archive-Orig-last-modified": "Thu, 03 Oct 1996 00:12:06 GMT"}, "memento-datetime": "Thu, 19 Dec 1996
08:24:28 GMT", "uri-m": "https://web.archive.org/web/19961219082428/http://www.sho.com:80/", "uri-r": 
"http://www.sho.com:80/"}

org,archive,web)/web/19961223174001/http:/www2.reference.com:80 {"created": "Wed, 12 Dec 2018 09:06:49 
GMT", "hash": "md5:b066243baa7a81d414c3e53b8a4982ef sha256:5b2efc24da6357062c2d9879b0e333359d20e7968244
f73801a599847f075b08", "hash-constructor": "(curl -s '$uri-m' && echo -n '$Content-Type $X-Archive-Orig
-date $X-Archive-Orig-etag $X-Archive-Orig-last-modified') | tee >(sha256sum) >(md5sum) >/dev/null | cut
-d ' ' -f 1 | paste -d':' <(echo -e 'md5\nsha256') - | paste -d' ' - -", "http-headers": {"Content-Type"
:"text/html", "Preference-Applied": "original-links, original-content", "X-Archive-Orig-date": "Mon, 23
Dec 1996 17:39:46 GMT", "X-Archive-Orig-etag": "\"bb98a-5a9-31ab9402\"", "X-Archive-Orig-last-modified": 
"Wed, 29 May 1996 00:02:10 GMT"}, "memento-datetime": "Mon, 23 Dec 1996 17:40:01 GMT", "uri-m": "https:
//web.archive.org/web/19961223174001/http://www2.reference.com:80/", "uri-r": "http://www2.reference.com
:80/"}
...
\end{Verbatim} 
 \caption{Block 1 contains 100 records (only four records are shown).}
\label{script:generate_block_ex}
\end{figure*}
\vspace{2mm}

{\raggedleft{}\textbf{disseminate\_block()}:} Pushes a block into two archives (\texttt{archive.}\\\texttt{org} and \texttt{perma.cc}). Again, because we are interested in calculating the average time of disseminating block, each block is pushed three time into both archives  resulting in creating 60 archived blocks (i.e., 30 per archive). We did not use \texttt{archive.is} and \texttt{webcitation.org} because .gz files were not handled correctly by those archives. Figure \ref{script:disseminate-block-example} shows an example of disseminating a block to two archives.

\begin{figure*}
\centering
%\userinput
%{\color{red}
\begin{Verbatim}[commandchars=\\\{\}]
$ python fixity.py \userinput{disseminate_block} http://manifest.ws-dl.cs.odu.edu/blocks



https://perma.cc/8YG3-X7KN

https://web.archive.org/web/20190121054059/https://manifest.ws-dl.cs.odu.edu/blocks/7bbf757046ac0a0a6001
5a1cb847c3189160d18c809b210073822df157609e01
\end{Verbatim}
%}
%\vspace{-0.5em}
\caption{An example of disseminating one block to two archives. The URI \texttt{http://manifest.ws-dl.cs.odu.edu/blocks} always redirects to the most recent published block. In this example the URI redirects to block 1: \texttt{https://manifest.ws-dl.cs.odu.edu/blocks/7bbf75704
6ac0a0a60015a1cb847c3189160d18c809b210073822df157609e01}.}
\label{script:disseminate-block-example}

\end{figure*}

\vspace{2mm}

{\raggedleft{}\textbf{verify\_block()}:} Accepts a URI-M, and discovers fixity information of the URI-M from the published blocks. Then, it computes current fixity information using \texttt{generate\_atomic()}. Finally, it compares current and discovered fixity. The function returns either ``Verified'' or ``Failed'' with other information, such as hash values. Figure \ref{script:verify-block-example} shows the output of \texttt{verify\_block()} for only 10 mementos (out of 1,000).

\begin{figure*}
\centering
%\userinput
%{\color{red}
\begin{Verbatim}[commandchars=\\\{\}]
\userinput{SN  Status  BlockIdx    TotalT  LookupT  GenerationT  VerifyT   URIM}
1  VERIFIED    1       0.82443  0.00144    0.82205    0.00073   https://web.archive.org/web/201512210445
08/http://www.justfocus.fr:80/arts/genie-leonard-de-vinci-pinacotheque.html
2  VERIFIED    1       0.53954  0.00157    0.53710    0.00063   https://web.archive.org/web/201509260235
12/http://evenium.fr:80/c/index
3  VERIFIED    1       0.59041  0.00145    0.58795    0.00076   https://web.archive.org/web/201511250509
59/http://centmillemilliards.com:80/wp/en/produit/humans-of-paris
4  VERIFIED    1       0.67178  0.00147    0.66912    0.00094   https://web.archive.org/web/201412080344
42/http://www.bergrettung-salzburg.at:80/home/
5  VERIFIED    1       0.59357  0.00169    0.59083    0.00081   https://web.archive.org/web/201403281737
22/http://carpalaid.com:80/
6  VERIFIED    1       0.80477  0.00161    0.80213    0.00078   https://web.archive.org/web/201410151107
04/http://www.airfrance.co.ao/cgi-bin/AF/AO/en/common/home/flights/ticket-plane.do
7  VERIFIED    1       0.67093  0.00176    0.66798    0.00093   https://web.archive.org/web/201707030456
56/https://www.abrasco.org.br/site/noticias/institucional/abrasco-convoca-comunidade-cientifica-para-a-m
archa-pela-ciencia/28036/
8  VERIFIED    1       0.82797  0.00154    0.82542    0.00078   https://web.archive.org/web/201707010935
44/https://sciencepop.fr/2017/04/13/climat-vaccin-ogm-francais-acceptent-science/
9  VERIFIED    1       0.76563  0.00156    0.76283    0.00099   https://web.archive.org/web/201702240949
32/http://www.damemagazine.com/2017/01/26/my-time-standing-rock-taught-me-what-we-need-do-resist
10 VERIFIED    1       0.79252  0.00172    0.78977    0.00078   https://web.archive.org/web/201605091517
44/http://www.ibc.ca:80/ab/disaster/fortmacfire
...
\end{Verbatim}
%}
%\vspace{-0.5em}
\caption{The output of \texttt{verify\_block()} (only the results of verifying 10 mementos out of 1,000 are shown). The column \textit{Status} indicates whether the fixity of a memento is verified or not. The column \textit{BlockIdx} is the block number used to verify the memento. The columns \textit{LookupT}, \textit{GenerationT}, and \textit{VerifyT} show the time taken for lookup the fixity information in blocks, generating current fixity information, and verifying/comparing the current fixity with the discovered fixity information from blocks, respectively. The column \textit{TotalT} shows the overall time taken to verify the fixity of the memento.}
\label{script:verify-block-example}

\end{figure*}

In addition to the Python scripts, we implemented the Archival Fixity server that is responsible for publishing and discovering manifests and blocks. For example, Figure \ref{fig:discover_manif} shows a request for discovering the closest manifest's creation date to December 22, 2018 for the given memento. The server response indicates that the closest manifest was created on December 12, 2018. 

\begin{figure*}
\centering 
     %\begin{minipage}[b]{8.5 cm}
     \begin{Verbatim}[commandchars=\\\{\}]
$ curl -I https://manifest.ws-dl.cs.odu.edu/manifest/\userinput{20181222}/https://web.archive.org/web/20171115140705
/http://rln.fm/

HTTP/2 302 Found

content-length: 501

content-type: text/html; charset=utf-8

date: Thu, 10 Jan 2019 09:16:40 GMT

\userinput{location}: https://manifest.ws-dl.cs.odu.edu/manifest/\userinput{20181212074423}/bd669de8835e38d54651fe9d04709515beec
          0c727db82a5366f4bc2506e103d8/https://web.archive.org/web/20171115140705/http://rln.fm/

server: ArchivalFixity/0.1
\end{Verbatim}
       % \vspace{-1.2em}
        
 \caption{Discovering the closest manifest to December 22, 2018 for the memento web. archive. org/web/2017111 5140705/http://rln.fm/.}
 %\vspace{-1.2em}
\label{fig:discover_manif}
\end{figure*}

%\begin{figure*}
%  \includegraphics[width=\textwidth]{images/redirect_server}
%  \caption{An example of discovering the closest manifest to December 22, 2018 for the memento https://web. archive. org/web/2017111 5140705/http://rln.fm/.}
%  \label{fig:discover_manif}
%\end{figure*}

\begin{figure}
\centering
%\setlength{\fboxsep}{0pt}%
%\fbox{
\includegraphics[width=235px]{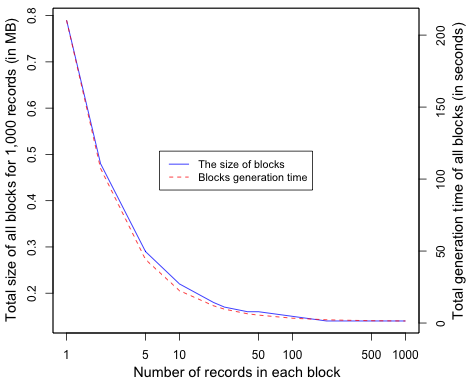}
%}
%\vspace*{-3mm}
\caption{The effect of the selected number of records per block.}
%\vspace*{-1mm}
\label{img:blocksst}
\end{figure}

The selected number of records per block affects the total size of all blocks and the time required to generate these blocks. Figure \ref{img:blocksst} illustrates that creating large blocks with a slowly growing chain is more efficient than a rapidly growing chain of small blocks. As mentioned in Section \ref{sec:verifying}, one factor of choosing the optimal number of records in each block is the largest size of a single block that can easily be stored in web archives. For example, we tested the Internet Archive (IA) to identify the largest single file that the archive can accept for preservation. After submitting multiple files with different sizes, we found that IA can accept up to 800 MB, beyond that the archive returns ``504 Gateway Time-out''.

\section{Results}
\begin{figure}
\centering
%\setlength{\fboxsep}{0pt}%
%\fbox{
\includegraphics[width=230px]{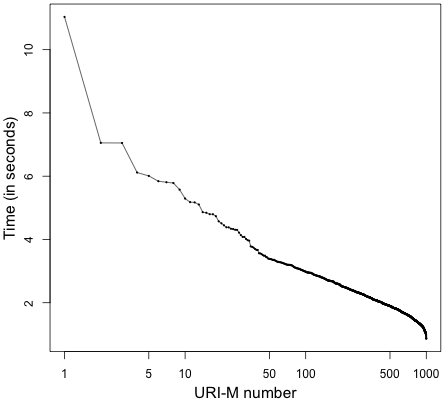}
%}
%\vspace*{-3mm}
\caption{Generating manifests of mementos.}
%\vspace*{-6mm}
\label{img:generate_manifetest}
\end{figure}

Figure \ref{img:generate_manifetest} illustrates the distribution of the average time taken to generate manifests. We generated three manifests for each memento, and calculated the average time, so the the total number of generated manifests is 3,000. The manifest generation time includes: 1) downloading the raw HTML content using the Requests module in Python, 2) calculating fixity information of the downloaded content, and 3) storing the fixity information locally in JSON format. The average size of the generated manifest files is 1,157 bytes. This size represents 2.79\% of the actual download HTML content, which is 41,392 bytes on average. The total size of all manifests is 1,156,657 bytes, while the total size of the blocks is 176,128 bytes. This indicates that the \emph{Block} approach requires less storage space than the \emph{Atomic} approach to store fixity information of the same number of mementos.   

\begin{figure}
\centering
%\setlength{\fboxsep}{0pt}%
%\fbox{
\includegraphics[width=230px]{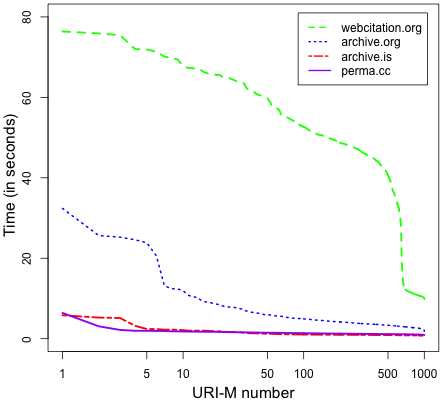}
%}
%\vspace*{-3mm}
\caption{Disseminating manifests to four archives.}
%\vspace*{-4mm}
\label{img:atomic_disseminate}
\end{figure}

\begin{figure}
\centering
%\setlength{\fboxsep}{0pt}%
%\fbox{
\includegraphics[width=230px]{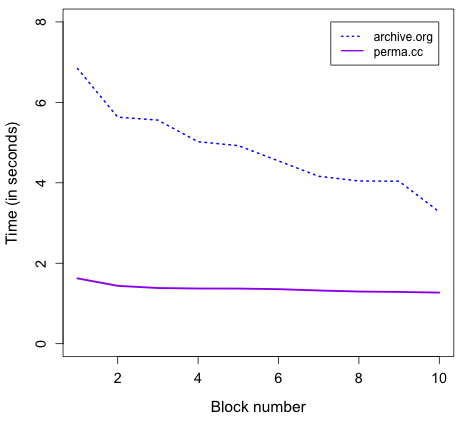}
%}
%\vspace*{-3.5mm}
\caption{Disseminating blocks to two archives.}
%\vspace*{-5mm}
\label{img:block_disseminate}
\end{figure}

As expected, the time for disseminating manifests and blocks was the maximum time compared with other operations, such as generating and verifying manifests. Figure \ref{img:atomic_disseminate} shows that pushing manifests into \texttt{webcitation.org} (or WebCite) takes much longer time than other archives. On average, we wait for 33.82 seconds before WebCite finishes processing an archival request of a manifest, while the manifest disseminating average time drops down dramatically in the other three archives as Table \ref{tab:time_averages} indicates. We observed that \texttt{archive.org} and \texttt{webcitation.org} add a few seconds response delay after receiving the first tens of archiving requests. In sum, it takes about 1.25X, 4X, and 36X longer to disseminate a manifest to \texttt{perma.cc}, \texttt{archive.org}, and \texttt{webcitation.org}, respectively, than \texttt{archive.is}, while it takes 3.5X longer to disseminate a block to \texttt{archive.org} than \texttt{perma.cc}. The average dissemination time of blocks in \texttt{archive.org} and \texttt{perma.cc} is shown in Figure \ref{img:block_disseminate}. 

Given a collection of $N$ mementos and $K$ web archives, the total number of resources that we are creating in the $K$ archives by the \emph{Atomic} and \emph{Block} approaches are $(N * K)$ and $(k*(N/B))$ respectively, where $B$ is the selected block size. In our study, $N=1,000$, $k_\text{atomic}=4$, $k_\text{block}=2$, and $B=100$. Then a total of $12,000$ resources were created by the \emph{Atomic} approach and only $60$ resources were created by the \emph{Block} approach considering the fact that we repeated the dissemination process for three times.

%\begin{figure*}[ht!]
 %   \centering
  %  \subfloat[Disseminating the blocks landing page to four archives.]%{{\includegraphics[height=220px]{images/block_disseminating_index} }} 
%    \subfloat[Disseminating the compressed blocks to two archives.]%{{\includegraphics[height=220px]{images/block_disseminating_ors_gz} }}\\[-2ex] 
%    \caption{The time taken  by the \emph{Block} approach for dissemination blocks}%
  %  \label{img:block_disseminate}%
%\end{figure*}

\begin{table}
\small
\centering
\caption {Average time (in seconds) for disseminating and downloading of manifests and blocks.}
%\vspace*{-2mm}
\setlength\tabcolsep{2.7pt}
\begin{tabular}{lcccc}
\hline
\textbf{Operation} & \textbf{archive.is} & \textbf{perma.cc} & \textbf{IA} & \textbf{WebCite} \\ \hline 
Manifest dissemination & 0.94 & 1.18 & 3.74 & 33.82  \\ \hline
Block dissemination & - & 1.37 & 4.80 & - \\ \hline
Manifest download & 0.47 & 0.60 & 1.42 & 4.55 \\ \hline
Block download & - & 0.30 & 7.19 & - \\ \hline
\end{tabular} 
%\vspace*{-3mm}
\label{tab:time_averages}
\end{table}

\begin{figure}
\centering
%\setlength{\fboxsep}{0pt}%
%\fbox{
\includegraphics[width=230px]{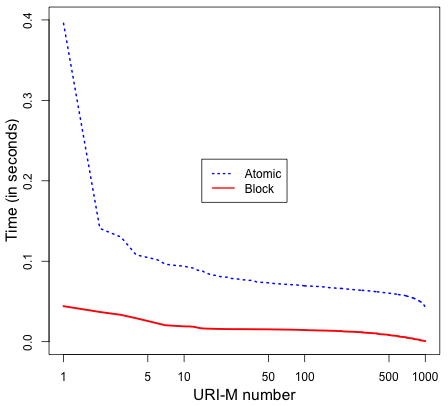}
%}
%\vspace*{-3mm}
\caption{Discovering manifests by both approaches.}
%\vspace*{-3mm}
\label{img:atomic_discov}
\end{figure}

\begin{figure}
\centering
%\setlength{\fboxsep}{0pt}%
%\fbox{
\includegraphics[width=230px]{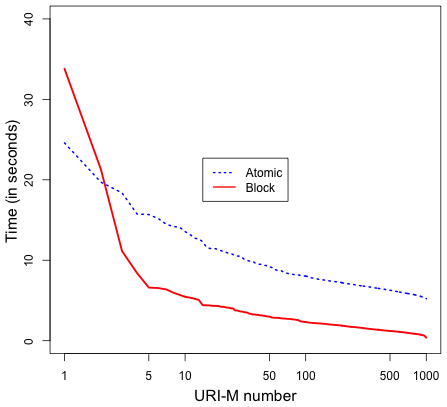}
%}
%\vspace*{-3mm}
\caption{Verifying mementos by both approaches.}
%\vspace*{-5mm}
\label{img:atomic_verify}
\end{figure}

Figure \ref{img:atomic_discov} shows the time required to discover manifests of each memento from the Archival Fixity server. Figure \ref{img:atomic_verify} illustates the total time for verifying the fixity of all mementos by both approaches. The verification time includes discovering manifests, computing current fixity information, downloading copies of manifests (in the \emph{Atomic} approach), and comparing manifests. On average, the verification time of a memento is 6.65 seconds by the \emph{Atomic} approach and 1.49 seconds by the \emph{Block} approach,  so the \emph{Block} approach performs 4.46X faster than the \emph{Atomic} approach on verifying the fixity of memento. Although we have predicted that some mementos might not be verified for reasons like an archive responds with ``HTTP 500 Error'', we have not yet encountered any failed cases (i.e., all mementos are verified successfully).

\section{Conclusions}
Most web archives do not allow users to access fixity information. Even if fixity information is accessible, it is provided by the same archive delivering content. In this proposal, we have described two approaches, \emph{Atomic} and \emph{Block}, for generating and verifying fixity of archived web pages. The proposed work does not require any change in the infrastructure of web archives and is built based on well-known standards, such as the Memento protocol. While a central service is used to create manifests, this approach does not exclude additional, centralized manifest servers, possibly tailored to specific communities. The \emph{Block} approach creates fewer resources in archives and reduces fixity verification time, while the \emph{Atomic} approach has the ability to verify fixity of archived pages even without involving the Archival Fixity server. On average, it takes about 1.25X, 4X, and 36X longer to disseminate a manifest to \texttt{perma.cc}, \texttt{archive.org}, and \texttt{webcitation.org}, respectively, than \texttt{archive.is}, while it takes 3.5X longer to disseminate a block to \texttt{archive.org} than \texttt{perma.cc}. The \emph{Block} approach performs 4.46X faster than the \emph{Atomic} approach on verifying the fixity of archived pages.

We believe that the \emph{Atomic} and \emph{Block} approaches can be adopted to verify fixity of particular archived web pages with important content. Some future improvements can be applied to those approaches so they become scalable and can work with any number of mementos. Varying or increasing the block size in the \emph{Block} approach might be one potential solution to improve its performance and reduce number of resources created in archives. Caching archived manifests in the Archival Fixity server should also improve the performance of the two approaches, so instead of discovering those manifests from the archives, we may used cached copies in the Archival Fixity server.

\section{Acknowledgements}
This work is supported in part by The Andrew W. Mellon Foundation (AMF) grant 11600663. This work includes contributions from Herbert Van de Sompel (DANS) and Martin Klein (LANL). We thank Ben Steinberg from the Perma.cc web archive for the generous increase in our monthly free quota to allow experimental fixity resource dissemination. We thank the WebCite archive for solving some technical issues about disseminating web resources.  

\clearpage

\bibliographystyle{ACM-Reference-Format}
\bibliography{ref} 

\end{document}